\documentclass[a4paper,fleqn]{cas-dc}
\usepackage[numbers]{natbib}
\usepackage{graphicx}
\usepackage{amsfonts}
\usepackage{amssymb}
\usepackage{amsmath}
\usepackage{accents}
\usepackage{placeins}
\usepackage{xfrac}
\usepackage{hyperref}

\newcommand\widearc[1]{\stackrel{\vbox{\offinterlineskip\hbox{\scriptsize$\frown$}\vskip-.55ex}}{#1}}
\newcommand\thickbar[1]{\widearc{#1}}
\newcommand{\be}{\begin{equation}}
\newcommand{\ee}{\end{equation}}
\newcommand{\bea}{\begin{eqnarray}}
\newcommand{\eea}{\end{eqnarray}}
\newcommand{\Eq}[1]{Eq.\,(\ref{#1})}
\newcommand{\Eqs}[1]{Eqs.\,(\ref{#1})}
\newcommand{\Fig}[1]{Fig.\,\ref{#1}}
\newcommand{\Sec}[1]{Section \ref{#1}}
\newcommand{\App}[1]{\autoref{#1}}
\newcommand{\br}{r}

\begin{document}

\let\WriteBookmarks\relax
\def\floatpagepagefraction{1}
\def\textpagefraction{.001}

\shorttitle{Foldy-Wouthuysen Green's function and WKB method for Dirac tunneling}
\shortauthors{M. B. Doost}
\title [mode = title]{Foldy-Wouthuysen Green's function and WKB transfer matrix method for Dirac tunneling through monolayer graphene with a mass gap}

\author[1]{Mark Behzad Doost}[orcid=0000-0003-4682-2889]
\address[1]{Independent Researcher, Hullbridge, Essex, England}
\ead{doostmb@gmail.com, https://gofund.me/34294631}
\cormark[1]
\cortext[cor1]{Corresponding author}
\begin{abstract}
I provide a transfer matrix method for the Foldy-Wouthuysen representation of the Dirac equation. I derive the relationship between the reflection and transmission coefficients of the Dirac spinors and the wavefunction in the transformed representation. I develop a WKB approximation for Dirac fermions that has the same elegant form as the WKB solution to Schr\"{o}dinger's equation. My WKB approximation is to all orders and includes the semi-classical turning point. I provide an extension to fully $2$-dimensional periodic structures by Fourier methods for band-gap engineering. I verify my methods for all energies by comparison with analytic solutions developed in the Dirac spinor representation. Rich appendices detail my research into the Green’s functions of Dirac fermions, where I rigorously derive the free space Green’s functions for the Foldy-Wouthuysen representation of the Dirac equation.%
\end{abstract}

%
%
%
\maketitle
\section{Introduction}
In Refs.~\cite{Barbie2010,PhysRevB.77.115446,Arovas2010} the pseudo-relativistic dispersion of $2$-dimensional ($2$D) materials was investigated theoretically for the emergence of extra Dirac points with $1$D periodic barriers. The method employed was the application of Bloch's theorem to transfer matrices of Dirac spinors. These band-gap engineering studies have attracted considerable interest since the vanishing band gap of $2$D materials presents a significant challenge for practical applications. Other methods for band-gap engineering used to create $2$D semiconductor materials include strain engineering \cite{Guinea2010}, vertical stacking \cite{Novoselov2016}, and chemical doping \cite{Wan2015}.

With the necessary knowledge for creating semiconductor $2$D materials, a theoretical study of the operational window for negative differential resistance was pursued by other authors \cite{Song2013}. The mathematical method employed in their study was the application of transfer matrices to the Dirac spinors of a rotated Dirac Hamiltonian derived in Ref.~\cite{Sonin2009}.

Motivated by the study of unconventional transmission and density of states, the modeling of Dirac fermions has advanced beyond idealized square barrier systems. In experimental devices, barriers are more often smoothly varying; semi-classical (WKB) methods are needed. An approach using separate Hamiltonians for electrons and holes in Ref.~\cite{Tudorovskiy2012} gave way in Ref.~\cite{Zalipaev2015} to an expansion in powers of $\hbar$ partly resembling the WKB method for the Schr\"{o}dinger equation. However, it is acknowledged by the authors of Refs.~\cite{Tudorovskiy2012,Zalipaev2015} that their mathematical approaches to semi-classical analysis can be divergent at the classical turning point.

This article contributes to the theory of tunneling in $2$D materials by applying the Foldy-Wouthuysen (FW) representation of the Dirac equation.
%

While the FW transformation was first conceived as a semi-relativistic representation of the Dirac equation \cite{Foldy1950}, I will show, through my rigorous derivations in \Sec{Sec:RSE1} through \Sec{Sec:RSE12} and my analytic and numerical verification of \Sec{Sec:RSE13} and \App{App:A2}, that the FW representation remains an exactly accurate representation of the Dirac equation for all energies. 

I find that the FW representation allows a derivation of the relativistic WKB approximation, \Sec{Sec:RSE6} to \Sec{Sec:RSE8b}, in the same elegant form as the approximate WKB solution to Schr\"{o}dinger's equation. Since at low energies, the FW representation reduces to the Schr\"{o}dinger equation, I have found, in \Sec{Sec:RSE9}, a simple method to mathematically describe the connections between WKB regions.

Fourier analysis is used in \Sec{Sec:RSE10} to extend my approach to fully $2$D periodic structures. \Sec{Sec:RSE10} offers the prospect of extending the scope of the band-gap engineering studies made in Refs.~\cite{Barbie2010,PhysRevB.77.115446,Arovas2010}. My numerical and analytic results showing that the FW equation is an exactly accurate representation of relativistic fermions, for all energies, will be important here. The Fourier transform method in \Sec{Sec:RSE10} requires expansion of the wavefunction in plane waves of momentum beyond semi-relativistic energies.

In \App{App:A4} I make the reader aware of a limit to appealing for solutions of the Dirac equation from its FW representation, I reveal and discuss a Green’s function (GF) paradox arising when transforming between the two representations. In spite of the paradox, I have rigorously derived the $1$D, $2$D and $3$D free space FW GFs in \App{App:A3} and \App{App:A5}. The Dyson equations for the FW GFs are available for all types of Born approximation \cite{Doost2015,Doost2016}.  

My article is organized as follows:
\Sec{Sec:RSE1} outlines the derivation of the FW representation from the Dirac equation.
\Sec{Sec:RSE2} gives the transmission of the Dirac equation in terms of the transmission of its FW representation. 
\Sec{Sec:RSE3} derives the boundary conditions of the FW equation at sharp steps in potential. 
\Sec{Sec:RSE4}  and \Sec{Sec:RSE5} develop the transfer matrices for rectangular barriers and delta potentials.
\Sec{Sec:RSE6}, \Sec{Sec:RSE7}, \Sec{Sec:RSE8} and  \Sec{Sec:RSE8b} give a $1st$, iterative and $2nd$ order WKB approximations for two interpretations of the FW equation.
\Sec{Sec:RSE9} gives the connecting formulae between regions where the WKB approximation is appropriate.
\Sec{Sec:RSE10} extends my approach to include periodic structures.
\Sec{Sec:RSE12} details how I evaluated the boundary conditions.
\Sec{Sec:RSE13} gives numerical results for examples of tunneling through barriers and resonant diodes.

The appendices are organized as follows:
\App{App:A1} analytically justifies my commutation of FW wavefunction operators. 
\App{App:A2} analytically calculates the reflection at a step in both the Dirac spinor and FW representations for comparison.
\App{App:A3a} examines the tunneling of a massless fermion through a magnetic delta barrier in the FW representation.
\App{App:A3} calculates $1$D GFs of the FW equation.
\App{App:A4} uncovers a relativistic GF paradox.
\App{App:A5} gives the correct $2$D and $3$D free space analytic FW GFs.
\section{Dirac equation and its Foldy-Wouthuysen representation}
\label{Sec:RSE1}
The Dirac equation for a scalar potential $V$ is given in Hamiltonian form by
\be\label{DIRAC}
\left[c{\bf\alpha}\cdot\hat{p}+\beta mc^2\right]\Psi=H\Psi=\left(E+V\right)\Psi\,,
\ee
\be
\gamma_t=\begin{pmatrix}
    {+I}&0\\
    0&{-I}
  \end{pmatrix}\,,
  \qquad
  \gamma_{x,y,z}=\begin{pmatrix}
    0&{+\sigma_{x,y,z}}\\
    {-\sigma_{x,y,z}}&0
  \end{pmatrix}\,,
\ee
\be
I=\begin{pmatrix}
    {+1}&0\\
    0&{+1}
  \end{pmatrix}\,,
  \qquad
  \sigma_{x}=\begin{pmatrix}
    0&{+1}\\
    {+1}&0
  \end{pmatrix}\,,
\ee
\be
\sigma_{y}=\begin{pmatrix}
    0&{-i}\\
    {+i}&0
  \end{pmatrix}\,,
  \qquad
  \sigma_{z}=\begin{pmatrix}
    {+1}&0\\
    0&{-1}
  \end{pmatrix}\,,
\ee
with ${\alpha_{x,y,z}}=\gamma_t\gamma_{x,y,z}$ and $\beta=\gamma_t$.
\be
\hat{p}=-i\hbar{\bf \nabla}=-i\hbar\left[\dfrac{\partial}{\partial x},\dfrac{\partial }{\partial y},\dfrac{\partial }{\partial z}\right]
\ee
is the momentum operator for the electron which is of mass $m$ and energy $E$. The charge and current density are given by the following well-known expressions:
\be\label{charged}
\rho=\Psi^{*}\Psi\,,
\ee
\be
j=\Psi^{*}\alpha\Psi\,.
\ee
There are two linearly independent solutions of the free particle Dirac equation,
\be\label{MD32}
\Psi^{\left(\pm\right)}_{\uparrow}=\begin{pmatrix}\mp mc^2-E-V \\ 0 \\ 0 \\ c\left(p_x\pm ip_y\right)\end{pmatrix}\exp{\left(\dfrac{ip_xx+ip_yy}{\hbar}\right)}\,,
\ee
\be\label{MD33}
\Psi^{\left(\pm\right)}_{\downarrow}=\begin{pmatrix} 0 \\\mp mc^2-E-V \\ c\left(p_x\mp ip_y\right)\\ 0 \end{pmatrix}\exp{\left(\dfrac{ip_xx+ip_yy}{\hbar}\right)}\,,
\ee
each can be normalized to one electron per unit volume by \Eq{charged}.

By making use of the FW transformation, I will decouple the differential
equations of the Dirac equation to give four independent identical equations.
I will show that by calculating the propagation for only one of these components I may obtain the full GF and transmission.

Under unitary transformation the Dirac equation becomes
\be
e^{+iS}He^{-iS}e^{+iS}\Psi=\left(E+V\right)e^{+iS}\Psi\,.
\ee
Foldy and Wouthuysen \cite{Foldy1950} found that their rotation
\be\label{eq:Markus}
e^{\pm iS}=\cos\theta\pm{\bf\beta}{\bf\alpha}\cdot\dfrac{\hat{p}}{\left|\hat{p}\right|}\sin\theta\,,
\ee
transformed the Dirac Hamiltonian into diagonal and anti-diagonal parts
\begin{multline}\label{Foundation}
\beta\left(mc^2\cos{\left(2\theta\right)}+c\left|\hat{p}\right|\sin{\left(2\theta\right)}\right)
\\
+c\alpha\cdot\hat{p}\left(\cos{\left(2\theta\right)}-\dfrac{mc}{\left|\hat{p}\right|}\sin{\left(2\theta\right)}\right)
\\
=e^{+iS}He^{-iS}\,.
\end{multline}
Subsequently it was chosen \cite{Foldy1950}
\be\label{ONEAA}
\cos\left(2\theta\right)=\dfrac{mc^2}{\sqrt{c^2\hat{p}^2+m^2c^4}}\,,
\ee
\be\label{THREEAA}
\sin\left(2\theta\right)=\dfrac{c\left|\hat{p}\right|}{\sqrt{c^2\hat{p}^2+m^2c^4}}\,,
\ee
so that by direct substitution of \Eq{ONEAA} and \Eq{THREEAA} into \Eq{Foundation}
\be\label{DONE}
\left[{\bf \beta}\sqrt{\hat{p}^2c^2+m^2c^4}-\hat{1}\left(V+E\right)\right]\widetilde{\Psi}=0\,.
\ee
\Eq{DONE} shows there are two possibilities for the FW equation
\be\label{OurFWEquation}
\left[\mp\sqrt{\hat{p}^2c^2+m^2c^4}-\left(V+E\right)\right]\widetilde{\Psi}^{\left(\pm\right)}=0\,,
\ee
$\widetilde{\Psi}^{\left(-\right)}$ corresponds to the electron, $\widetilde{\Psi}^{\left(+\right)}$ to the positron. The rotated wavefunction $\widetilde{\Psi}$ is related to $\Psi$ by
\be\label{FWRotation}
\left[\cos\theta+{\bf\beta}{\bf\alpha}\cdot\dfrac{\hat{p}}{\left|\hat{p}\right|}\sin\theta\right]\Psi=\widetilde{\Psi}\,.
\ee
Taylor expanding $\sqrt{\hat{p}^2c^2+m^2c^4}$ in \Eq{OurFWEquation} gives
\be
\left[\mp\mathcal{L}-\left(V+E\pm mc^2\right)\right]\widetilde{\Psi}^{\left(\pm\right)}=0\,,
\label{FWEquation}
\ee
where
\be\label{FT43}
\mathcal{L}=mc^2\left[\dfrac{1}{2}\left[{\dfrac{\hat{p}}{mc}}\right]^2-\dfrac{1}{8}\left[{\dfrac{\hat{p}}{mc}}\right]^4+\dfrac{1}{16}\left[{\dfrac{\hat{p}}{mc}}\right]^6-...\right]\,.
\ee
\Eq{OurFWEquation} and \Eq{FWEquation} are two interpretations of the FW equation.

I demonstrate the unsuitability of my FW approach for calculating magnetic barrier tunneling in \App{App:A3a}. I rigorously derive the GFs of the FW \Eq{FWEquation} in \App{App:A3} and \App{App:A5}. I uncover and discuss a GF paradox of the FW \Eq{FWEquation} in \App{App:A4}. 
\section{Relationship between the transmission of the Dirac equation and its Foldy-Wouthuysen representation}
\label{Sec:RSE2}
Throughout my numerical and analytic results, \Sec{Sec:RSE13} and \App{App:A2}, I will calculate the reflection coefficient $R$ to deduce the transmission coefficient $T$ as
\be
T=1-R\,.
\ee
Calculating $R$, rather than $T$ directly, avoids the need for wavefunction normalization in both the incident and transmitted regions.

Let $\Psi$ be given by \Eq{MD32} and \Eq{MD33} as
\be\label{BCC3}
\Psi=u\exp{\left(\dfrac{ip_xx+ip_yy}{\hbar}\right)}\,,
\ee
taking $\widetilde{\Psi}$ of the form
\be\label{Fexppxpy}
\widetilde{\Psi}=\widetilde{u}\exp{\left(\dfrac{ip_xx+ip_yy}{\hbar}\right)}\,,
\ee
I find by application of \Eq{FWRotation} to \Eq{BCC3}
\be\label{FWtoDirac1}
\begin{split}
&\widetilde{u}_1=u_1\cos\theta+\left(d_x-id_y\right)u_4\sin\theta\,,\\
&\widetilde{u}_2=u_2\cos\theta+\left(d_x+id_y\right)u_3\sin\theta\,,\\
&\widetilde{u}_3=u_3\cos\theta-\left(d_x-id_y\right)u_2\sin\theta\,,\\
&\widetilde{u}_4=u_4\cos\theta-\left(d_x+id_y\right)u_1\sin\theta\,,
\end{split}
\ee
where $\hat{d}=\hat{p}/|\hat{p}|$.

\Eq{BCC3}, \Eq{Fexppxpy}, and \Eqs{FWtoDirac1} show $\widetilde{\Psi}$ in exponential planar form which is useful for calculating the fermion propagation through homogeneous space.

The incident ($i$) and reflected ($r$) waves for planar structures in the Dirac equation description are related to the reflection coefficient in the well-known way: 
\be\label{BeginTransFormulas}
R=\dfrac{\left(\Psi\cdot\Psi^*\right)^{\left(r\right)}}{\left(\Psi\cdot\Psi^*\right)^{\left(i\right)}}\,.
\ee
For the two linearly independent solutions of the Dirac equation, \Eq{MD32} and \Eq{MD33}, \Eq{BeginTransFormulas} gives rise to two degenerate expressions for the reflection coefficient:
\be\label{transmission0}
R_{\uparrow}=\dfrac{\left(\left|\Psi_1\right|^2+\left|\Psi_4\right|^2\right)^{(r)}}{\left(\left|\Psi_1\right|^2+\left|\Psi_4\right|^2\right)^{\left(i\right)}}\,,
\ee
\be\label{transmission1}
R_{\downarrow}=\dfrac{\left(\left|\Psi_2\right|^2+\left|\Psi_3\right|^2\right)^{\left(r\right)}}{\left(\left|\Psi_2\right|^2+\left|\Psi_3\right|^2\right)^{\left(i\right)}}\,.
\ee
I take note of \Eqs{FWtoDirac1} linking $\widetilde{u}$ to $u$ and calculate when $d$ is real
\be
\left|\widetilde{u}_1\right|^2+\left|\widetilde{u}_4\right|^2=\left|u_1\right|^2+\left|u_4\right|^2\,,
\ee
\be\label{EndTransFormulas}
\left|\widetilde{u}_2\right|^2+\left|\widetilde{u}_3\right|^2=\left|u_2\right|^2+\left|u_3\right|^2\,,
\ee
therefore $R$ described by the Dirac equation can be rewritten in terms of the FW wavefunctions:
\be\label{transmission2}
R_{\uparrow}=\dfrac{\left(\left|\widetilde{\Psi}_1\right|^2+\left|\widetilde{\Psi}_4\right|^2\right)^{\left(r\right)}}{\left(\left|\widetilde{\Psi}_1\right|^2+\left|\widetilde{\Psi}_4\right|^2\right)^{\left(i\right)}}\,,
\ee
\be\label{transmission3}
R_{\downarrow}=\dfrac{\left(\left|\widetilde{\Psi}_2\right|^2+\left|\widetilde{\Psi}_3\right|^2\right)^{\left(r\right)}}{\left(\left|\widetilde{\Psi}_2\right|^2+\left|\widetilde{\Psi}_3\right|^2\right)^{\left(i\right)}}\,.
\ee

Since the components of $\widetilde{\Psi}$ are degenerate I am only required to evaluate the propagation of a fermion by the FW \Eq{FWEquation} once when calculating $R$ and $T$.
\section{Boundary Conditions of the Foldy-Wouthuysen equation at a sharp step}
\label{Sec:RSE3}
Consider a sharp step in potential at $x=a$.
I will integrate the FW \Eq{FWEquation} through the boundary, normal to the boundary
\be
\lim_{\left({a_+}-{a_-}\right)\rightarrow 0}\int^{a_+}_{a_-}\left[\mp\mathcal{L}-\left(V+E\pm mc^2\right)\right]\widetilde{\Psi}^{\left(\pm\right)}\dfrac{dx}{c\hbar}=0\,.
\ee

In this section, I introduce the operator $\mathcal{\thickbar{L}}$, which commutes with $c\hat{p}_x$
and is defined by
\be\label{AppA}
\mathcal{L}\widetilde{\Psi}=-\mathcal{\thickbar{L}}c\hat{p}_x\widetilde{\Psi}=-c\hat{p}_x\mathcal{\thickbar{L}}\widetilde{\Psi}\,.
\ee
\App{App:A1} discusses \Eq{AppA}. 

Since
\be
\lim_{\left({a_+}-{a_-}\right)\rightarrow 0}\int^{a_+}_{a_-}\left(V+E\pm mc^2\right)\widetilde{\Psi}^{\left(\pm\right)}\dfrac{dx}{c\hbar}=0\,,
\ee
due to the function integrated being finite but the integration range tending to zero, I have
\be\label{Result}
\left[i\mathcal{\thickbar{L}}\widetilde{\Psi}\right]^{a_+}_{a_-}=\lim_{\left({a_+}-{a_-}\right)\rightarrow 0}\int^{a_+}_{a_-}\mathcal{L}\widetilde{\Psi}\dfrac{dx}{c\hbar}=0\,.
\ee
I can understand \Eq{Result} when I consider that
\be
\mathcal{L}\widetilde{\Psi}=ic\hbar\dfrac{\partial \left(\mathcal{\thickbar{L}}\widetilde{\Psi}\right)}{\partial x}\,.
\ee
\Eq{Result} shows that $\widetilde{\Psi}$ is continuous under the operation of $\mathcal{\thickbar{L}}$. 

In order to evaluate $\mathcal{\thickbar{L}}\widetilde{\Psi}$ for the mathematical description of the boundary conditions note
\be
\pm\mathcal{\thickbar{L}}c\hat{p}_x\widetilde{\Psi}^{\left(\pm\right)}=\left(V+E\pm mc^2\right)\widetilde{\Psi}^{\left(\pm\right)}=\pm cp_x\mathcal{\thickbar{L}}\widetilde{\Psi}^{\left(\pm\right)}\,,
\ee
therefore
\be
\mathcal{\thickbar{L}}\widetilde{\Psi}^{\left(\pm\right)}=\pm\dfrac{V+E\pm mc^2}{cp_x}\widetilde{\Psi}^{\left(\pm\right)}=\mp\alpha^{\left(\pm\right)}\widetilde{\Psi}^{\left(\pm\right)}\,.
\ee

For my second boundary condition I note that $\hat{p}\widetilde{\Psi}$ always gives a finite observed value, therefore $\widetilde{\Psi}$ is continuous everywhere.

My boundary conditions are in stark contrast to the continuity conditions of the Klein-Gordon equation; in that case, continuity is with respect to momentum and wavefunction. I verify my boundary conditions numerically and analytically in \Sec{Sec:RSE13} and \App{App:A2}.
\section{Transfer matrices for a rectangular potential barrier}
\label{Sec:RSE4}
Inside a layer thickness $D$ the transfer matrix of $\widetilde{\Psi}$ is given by
\be\label{FIFTY}
\begin{pmatrix}
    {\exp{\left(-\dfrac{iDp_x}{\hbar}\right)}}&0\\
    0&{\exp{\left(+\dfrac{iDp_x}{\hbar}\right)}}
  \end{pmatrix}\begingroup\renewcommand*{\arraystretch}{2.0}\begin{pmatrix} \widetilde{a}_1 \\ \widetilde{b}_1 \end{pmatrix}\endgroup=\begingroup\renewcommand*{\arraystretch}{2.0}\begin{pmatrix} \widetilde{a}_0 \\ \widetilde{b}_0 \end{pmatrix}\endgroup\,,
\ee
where $\widetilde{a}_0$ and $\widetilde{a}_1$ denote forward traveling FW wave parts and $\widetilde{b}_0$ and $\widetilde{b}_1$ denote backward travelling FW wave parts, 
$0$ denotes on the left, $1$ denotes on 
the right. The fermion is refracted at an angle $\phi$ to the layer boundary normal,
where
\be
p_x=p\cos\phi\,.
\ee

For a boundary at a sharp step, applying the wavefunction continuity condition set out in \Sec{Sec:RSE3} gives
\be\label{SMOOTH}
\widetilde{a}_0+\widetilde{b}_0=\widetilde{a}_1+\widetilde{b}_1\,.
\ee
Secondly, applying wavefunction continuity under the operation of $\mathcal{\thickbar{L}}$, derived in \Sec{Sec:RSE3}, to \Eq{SMOOTH} gives
\be\label{CONT}
\alpha_0\widetilde{a}_0-\alpha_0\widetilde{b}_0=\alpha_1\widetilde{a}_1-\alpha_1\widetilde{b}_1\,.
\ee
\Eq{SMOOTH} and \Eq{CONT} can be written in matrix form as
\begin{gather}
\dfrac{1}{2}\begin{pmatrix}
    {\left(1+\dfrac{\alpha_1}{\alpha_0}\right)}&{\left(1-\dfrac{\alpha_1}{\alpha_0}\right)}\\
    {\left(1-\dfrac{\alpha_1}{\alpha_0}\right)}&{\left(1+\dfrac{\alpha_1}{\alpha_0}\right)}
  \end{pmatrix}\begingroup\renewcommand*{\arraystretch}{2.0}\begin{pmatrix} \widetilde{a}_1 \\ \widetilde{b}_1 \end{pmatrix}\endgroup=\begingroup\renewcommand*{\arraystretch}{2.0}\begin{pmatrix} \widetilde{a}_0 \\ \widetilde{b}_0 \end{pmatrix}\endgroup\,.
\end{gather}

During tunneling, wave modes either decay or grow exponentially. Transfer matrix multiplication combines these large and small values, this process is numerically unstable and inaccurate. Instead, I suggest using the method of Refs.~\cite{Ko1988,PhysRevB.66.045102} to combine my transfer matrices. The method of Refs.~\cite{Ko1988,PhysRevB.66.045102} addresses these instability issues by separating the exponentially growing and decaying terms.
\section{Transfer matrix for a delta potential barrier}
\label{Sec:RSE5}
Integrating \Eq{FWEquation}, with $V(x)=g\delta(x)$ and with respect to $dx/c\hbar$, across $\left[0_{-},0_{+}\right]$ gives
\be\label{SMOOTH1}
\left[\mp i\mathcal{\thickbar{L}}\left(\widetilde{a}^{\left(\pm\right)}+\widetilde{b}^{\left(\pm\right)}\right)\right]^{0_+}_{0_-}=\dfrac{g}{c\hbar}\left(\widetilde{a}^{\left(\pm\right)}+\widetilde{b}^{\left(\pm\right)}\right)\biggr\rvert_{x=0}\,.
\ee
Continuity of the wavefunction gives
\be\label{SMOOTH2}
\widetilde{a}_0+\widetilde{b}_0=\widetilde{a}_1+\widetilde{b}_1\,.
\ee
\Eq{SMOOTH1} and \Eq{SMOOTH2} can be written in the following matrix form
\begin{gather}
\begin{pmatrix}
    {\left(1-\dfrac{ig}{2c\hbar\alpha_k}\right)}&{\left(-\dfrac{ig}{2c\hbar\alpha_k}\right)}\\
    {\left(+\dfrac{ig}{2c\hbar\alpha_k}\right)}&{\left(1+\dfrac{ig}{2c\hbar\alpha_k}\right)}
  \end{pmatrix}\begingroup\renewcommand*{\arraystretch}{2.0}\begin{pmatrix} \widetilde{a}_1 \\ \widetilde{b}_1 \end{pmatrix}\endgroup=\begingroup\renewcommand*{\arraystretch}{2.0}\begin{pmatrix} \widetilde{a}_0 \\ \widetilde{b}_0 \end{pmatrix}\endgroup\,,
\end{gather}
where I define $\widetilde{a}_0$, $\widetilde{a}_1$, $\widetilde{b}_0$, and $\widetilde{b}_1$ in the same way as \Sec{Sec:RSE4}.
\section{${\bf 1st}$ order relativistic WKB approximation}
\label{Sec:RSE6}
Consider the following result from \App{App:A3} derived from the correct fermion interpretation of the FW equation in terms of $\mathcal{L}$,
\be\label{EAMG}
T\left(k\right)=\dfrac{\left(2\alpha_k\alpha_q\right)^2}{\left(\alpha^2_q-\alpha^2_k\right)^2\sin^2\left(\dfrac{2m^*qa}{\hbar}\right)+\left(2\alpha_k\alpha_q\right)^2}
\ee
for a rectangular barrier given by 
\be
V\left(x\right)=\left\{
\begin{array}{lc}  V & \mbox{for}\ \  \left|x\right|<a\,,\\
0 & \mbox{for}\ \  \left|x\right|>a\,.\end{array}\right.
 \label{unpslab}
\ee
\Eq{EAMG} can be approximated as 
\be\label{EAMG2}
T\left(k\right)\approx\left(\dfrac{2\alpha_k\alpha_q}{\alpha^2_k-\alpha^2_q}\right)^2\exp{\left(-\dfrac{4im^*qa}{\hbar}\right)}
\ee
by assuming exponential wavefunction decay within the barrier. Taking logarithms of \Eq{EAMG2}, the non-logarithmic term dominates the logarithmic term so that
\be
\log_eT\left(k\right)\approx -\dfrac{4im^*qa}{\hbar}\,.
\ee

Assuming independence of transmission events, the total transmission probability for crossing an overall barrier is
\be\label{TransmissionPi}
T\approx\prod_{i=1}^{\infty}T_{i}\,,
\ee
\be\label{TransmissionLogSum}
\log_e\prod_{i=1}^{\infty}T_{i}\approx\sum_{i=1}^{\infty}-\dfrac{2im^*_iq_i\Delta x}{\hbar}=-\dfrac{2i}{\hbar}\int p_xdx\,,
\ee
where the $T_i$ are the transmission probabilities of the individual barriers. I see from \Eq{TransmissionLogSum} that inside a smoothly varying layer the FW wavefunction is approximately given by
\be\label{SimpleWKB}
\widetilde{\Psi}\approx\exp{\left({\pm\dfrac{i}{\hbar}\int p_xdx}\right)}\prod_{i=1}^{N}\left(\dfrac{2\alpha_k\alpha_q}{\alpha^2_k-\alpha^2_q}\right)_i\,.
\ee
\section{Iterative relativistic WKB approximation}
\label{Sec:RSE7}
Consider a wavefunction of the form
\be\label{GeneralFormOfWaveFunction}
\widetilde{\Psi}=\exp{\left(\dfrac{i\Phi}{\hbar}\right)}\,,
\ee
then the boson FW interpretation \Eq{OurFWEquation} becomes
\be\label{DONE_TWO}
\left[\sqrt{c^2\left(\dfrac{\partial\Phi}{\partial x}\right)^2-c^2i\hbar\dfrac{\partial^2\Phi}{\partial x^2}+m^2c^4}-\left(V+E\right)\right]\widetilde{\Psi}=0\,.
\ee
\Eq{DONE_TWO} can be exactly rewritten  as 
\be\label{DONE_THREE}
c^2\left(\dfrac{\partial\Phi}{\partial x}\right)^2-c^2i\hbar\dfrac{\partial^2\Phi}{\partial x^2}+m^2c^4=\left(V+E\right)^2\,.
\ee
I now make use of $(E+V)^2=m^2c^4+p^2c^2$ in \Eq{DONE_THREE} and find that as expected
\be\label{DONE_FOUR}
\hat{p}^2_x\widetilde{\Psi}=p^2_x\widetilde{\Psi}\,,
\qquad
\dfrac{\partial\Phi}{\partial x}=\sqrt{p^2_x+i\hbar\dfrac{\partial^2\Phi}{\partial x^2}}\,.
\ee
I know from \Sec{Sec:RSE6}
\be
\pm\dfrac{\partial\Phi}{\partial x}\approx p_x\,,
\ee
suggesting a next approximation on from \Eq{SimpleWKB} is
\be\label{DONE_FOUR2}
\Phi\approx\pm\int\sqrt{p^2_x\pm i\hbar\dfrac{\partial p_x}{\partial x}}dx\,. 
\ee
\Eq{DONE_FOUR2} follows from the integration of \Eq{DONE_FOUR}. This procedure can be iterated to an $n^{th}$ order WKB approximation in the boson FW interpretation. 
\section{${\bf 2nd}$ order relativistic WKB approximation}
\label{Sec:RSE8} 
Expanding $\Phi\left(x\right)$ in powers of $\hbar$
\be\label{SExpansion}
\Phi\left(x\right)=\Phi_0\left(x\right)+\hbar \Phi_1\left(x\right)+\hbar^2\Phi_2\left(x\right)+...\,,
\ee
then substituting \Eq{SExpansion} into \Eq{DONE_FOUR} gives
\be\label{TextBook}
\left(\Phi_0'+\hbar\Phi_1'+...\right)^2-i\hbar\left(\Phi_0''+\hbar\Phi_1''+...\right)=p^2_x\,.
\ee
Equating terms in powers of $\hbar$, in \Eq{TextBook}, I see 
\be
\pm\Phi_0'=p_x\,,
\qquad
2\Phi_0'\Phi_1'-i\Phi_0''=0\,,
\ee
so that
\be\label{MomentumFormS}
\Phi_0=\pm\int p_{x}dx\,,
\qquad
\Phi_1=\dfrac{i}{2}\log_e\Phi_0'\,.
\ee
\Eqs{MomentumFormS}, \Eq{SExpansion}, and \Eq{GeneralFormOfWaveFunction} give
\be\label{KleinGordonWKBSecondOrder}
\widetilde{\Psi}\approx \dfrac{\exp{\left(\pm \dfrac{i}{\hbar}\int p_{x}dx\right)}}{\sqrt{p_x}}\,. 
\ee
This result is identical to the Klein-Gordon WKB approximation because on the way I approximated FW \Eq{FWEquation} with \Eq{OurFWEquation}.
\section{Discussion of fermion and boson WKB approximations}
\label{Sec:RSE8b}

When I made my WKB approximation in \Sec{Sec:RSE6}, from what I will show in \Sec{Sec:RSE13} and \App{App:A2} to be the correct form \Eq{FWEquation} of the FW equation, I saw in the non-exponential factor of \Eq{EAMG2} appearances of $p_i$ changed to appearances of $\alpha_i$ going from boson to fermion descriptions. I have noticed, that for tunneling through slowly varying potentials, \Eq{EAMG2}, \Eq{TransmissionPi} and \Eq{KleinGordonWKBSecondOrder} give the implication:
\be\label{TransmissionPi2}
\prod_{i=1}^{N}\left(\dfrac{2kq}{k^2-q^2}\right)^2_i\approx \dfrac{p_1}{p_N}\implies \prod_{i=1}^{N}\left(\dfrac{2\alpha_k\alpha_q}{\alpha^2_k-\alpha^2_q}\right)^2_i\approx \dfrac{\alpha^{\left(\pm\right)}_{p_1}}{\alpha^{\left(\pm\right)}_{p_N}}\,.
\ee
\Eq{TransmissionPi2} suggests that the WKB approximation for fermions, in the correct FW representation, will take the form
\be\label{DiracWKBSecondOrder}
\widetilde{\Psi}\approx\dfrac{\exp{\left(\pm \dfrac{i}{\hbar}\int p_{x}dx\right)}}{\sqrt{\alpha_{p_x}}}\,.
\ee
Going from \Eq{KleinGordonWKBSecondOrder} to \Eq{DiracWKBSecondOrder} I swapped an appearance of $p_i$ to $\pm\alpha^{\left(\pm\right)}_{i}$, from boson to fermion descriptions, as suggested by \Eq{EAMG2}, \Eq{TransmissionPi} and \Eq{KleinGordonWKBSecondOrder}.
\section{Connection formulae}
\label{Sec:RSE9}
It is suggested by \Sec{Sec:RSE7} that for the WKB approximation to be appropriate
\be\label{ConditionWKB}
\left|p_x\right|^2\gg\left|\hbar\dfrac{\partial p_x}{\partial x}\right|\,.
\ee
\Eq{ConditionWKB} suggests that when $p_x\approx 0$ the WKB method fails. I need a connection formula between the regions where $p_x^2\neq 0$.

I assume that when \Eq{ConditionWKB} fails
\be
mc\gg\left|p_x\right|\,,
\ee
so that I may treat the connecting region with the $c=\infty$ non-relativistic limit of the FW \Eq{FWEquation}. The wavefunction in the connecting region is then given by the Airy function solutions to the Schr\"{o}dinger equation. This approach leads to the two connected solutions for $\widetilde{\Psi}$ between the classical and non-classical regions:

{\bf Classical region ${\bf p^2>0}$ and ${\bf x<x_1}$}
\be\label{Region1}
A_1\dfrac{\exp{\left(+\dfrac{i}{\hbar}\int^x_{x_1}\left|p\right|dx'\right)}}{\sqrt{\left|p\right|}}+B_1\dfrac{\exp{\left(-\dfrac{i}{\hbar}\int^x_{x_1}\left|p\right|dx'\right)}}{\sqrt{\left|p\right|}}\,,
\ee

{\bf Non-classical region ${\bf p^2<0}$ and ${\bf x>x_1}$}
\be\label{Region2}
A_2\dfrac{\exp{\left(-\dfrac{1}{\hbar}\int^x_{x_1}\left|p\right|dx'\right)}}{\sqrt{\left|p\right|}}+B_2\dfrac{\exp{\left(+\dfrac{1}{\hbar}\int^x_{x_1}\left|p\right|dx'\right)}}{\sqrt{\left|p\right|}}\,,
\ee
where
\be
2A_2=-i\left(A_1+iB_1\right)\exp\big(+i\pi/4\big)
\ee
and
\be
B_2=+i\left(A_1-iB_1\right)\exp\big(-i\pi/4\big)\,.
\ee
In \Eq{Region1} and \Eq{Region2} I have used the asymptotic forms of the Airy functions and taken $p=0$ to be at $x_1$. 

The wavefunctions \Eq{Region1} and \Eq{Region2} are in the WKB form derived for relativistic fermions in \Sec{Sec:RSE6} to \Sec{Sec:RSE8b}, therefore the solutions in both regions can be extended to relativistic energies.
\section{Extension to periodic structures}
\label{Sec:RSE10}
Consider the periodic potential
\be\label{VF}
V\left(x\right)=\sum_GV_G\exp{\left(+\dfrac{iGx}{\hbar}\right)}\,.
\ee
\Eq{VF} is a Fourier expansion of the potential, where $G=2\pi n/a$ and $n$ is positive and negative integers. The wavefunction follows from Bloch's theorem for periodic crystals
\be\label{PF}
\widetilde{\Psi}^{\left(\pm\right)}=\exp{\left(+\dfrac{ikx}{\hbar}\right)}\sum_gA_g\exp{\left(-\dfrac{igx}{\hbar}\right)}\,,
\ee
where $g=2\pi m/a$ and $m$ is positive and negative integers. 

Substituting \Eq{VF} and \Eq{PF} into
\be\label{Problem}
\left[\mp\mathcal{L}-\left(E\pm mc^2+V\right)\right]\widetilde{\Psi}^{\left(\pm\right)}=0
\ee
and multiplying through by $\exp{(i(G'-k)x/\hbar)}$, where $G'=2\pi n'/a$, then integrating $x$ over one period $a$, I find that the first, second, and third terms are only non-zero for $g=G'$ while the fourth term is only non-zero for $g-G=G'$. Therefore, with rearrangement
\be\label{2DSuperLattice}
\sum_g\left(\delta_{gG'}W_g-V_{g-G'}\right)A_g=\left(E\pm mc^2\right)A_{G'}\,,
\ee
where
\be
W_g=-c\left(k-g\right)\alpha^{(\pm)}_{k-g}\,.
\ee

The eigenvalue problem described by \Eq{2DSuperLattice} can be solved for varying $k$ to give the eigenvalues $\left(E\pm mc^2\right)$. The relationship between $k$ and $E$ defines the band structure.

For planar structures of period $D$ I have by Bloch's theorem 
\be
\widetilde{\Psi}\left(x+D\right)=\exp{\left(\dfrac{iKD}{\hbar}\right)}\widetilde{\Psi}\left(x\right)\,,
\ee
which in terms of transfer matrices can be written as
\be\label{MM2}
M\widetilde{\Psi}=\exp{\left(\dfrac{iKD}{\hbar}\right)}\widetilde{\Psi}=\lambda\widetilde{\Psi}\implies\left|M-\lambda I\right|=0\,, 
\ee
where
\be
M=\begin{pmatrix}
    {M_{11}}&{M_{12}}\\
    {M_{21}}&{M_{22}}
  \end{pmatrix}\,.
\ee
For fermions to propagate through the periodic structure, $K$ must be real. Consequently, by \Eqs{MM2}
\be\label{BEATS}
-1\leq\cos\left(\dfrac{KD}{\hbar}\right)=\dfrac{M_{11}+M_{22}}{2}\leq +1\,.
\ee
Band gaps arise in $1$D periodic potentials if the condition in \Eq{BEATS} is not satisfied.
\section{Calculation of transfer matrix parameters}
\label{Sec:RSE12}
\subsection{Calculation of the angle of refraction}
To match the Dirac spinors of \Eq{MD32} and \Eq{MD33} either side of a potential step at $x=0$ it must be that
\be
p_y^{\left(i\right)}|_{x=0}=p_y^{\left(t\right)}|_{x=0}\,,
\ee
therefore, since
\be
\hat{p}={\hat{x}}p\cos\phi+{\hat{y}}p\sin\phi\,,
\ee
I arrive at Snell's law for the relativistic fermion
\be
\dfrac{\sin\phi^{\left(i\right)}}{\sin\phi^{\left(t\right)}}=\dfrac{p^{\left(t\right)}}{p^{\left(i\right)}}\,.
\ee
I do not analytically verify non-normal incident relativistic tunneling and transmission in this article.
\subsection{Tunneling between layers of varying electron mass}
In free space by special relativity
\be
E^2=m^2c^4\left(1+\left(\dfrac{k}{c}\right)^2\right)\,,
\label{CCCA}
\ee
inside a layer by special relativity
\be
\left(E+V\right)^2={m^*}^2{c^*}^4\left(1+\left(\dfrac{q}{c^*}\right)^2\right)\,,
\label{CCCB}
\ee
where ${m^{*}}$ is the effective mass of the electron in the layer and $c^{*}$ is the corresponding effective speed of light. Comparing \Eq{CCCA} and \Eq{CCCB}, I find
\begin{multline}\label{WaveNumber}
\dfrac{q^2}{k^2}=\left(\dfrac{mc}{m^*c^*}\right)^2+\dfrac{mc^2}{m^*{c^*}^2}\dfrac{2g(k/c)V}{m^{*}k^2}+\left(\dfrac{V}{m^*c^*k}\right)^2+
\\
+\dfrac{\left(mc^2-m^*{c^*}^2\right)\left(mc^2+m^*{c^*}^2\right)}{\left(m^{*}c^*\right)^2}\dfrac{1}{k^2}\,.
\end{multline}
In the last term of \Eq{WaveNumber} I used the difference of two squares to improve numerical accuracy. When $|k|<c$ the function $g(k/c)$
is given by the Taylor expansion
\be\label{gggg1}
g\left(k/c\right)=\sqrt{1+\left(\dfrac{k}{c}\right)^2}=1+\dfrac{\left(k/c\right)^2}{2}-\dfrac{\left(k/c\right)^4}{8}+...
\ee
and when $|k|>c$ the function $g(k/c)$ is given by the Taylor expansion
\be\label{gggg2}
g\left(k/c\right)=\left(\dfrac{k}{c}\right)\left(1+\dfrac{\left(c/k\right)^2}{2}-\dfrac{\left(c/k\right)^4}{8}+...\right)\,. \ee
Computationally, I add the smallest terms together first to improve numerical accuracy. In terms of $g(k/c)$ and the $x$-component of electron momentum $m^*q_x$ I calculate the parameter $\alpha^{(\pm)}_q$ to be
\be
\alpha^{\left(\pm\right)}_q=-\dfrac{mc^2g\left(k/c\right)+V\pm m^*{c^*}^2}{c^*m^*q_x}\,.
\ee
\subsection{Tunneling into a layer where electron mass vanishes}
In free space the fermion momentum dispersion is given by \Eq{CCCA} whereas inside the layer electron mass has vanished and therefore
\be
E+V=qc^*\,,
\label{NoMassCCCB}
\ee
where $q$ is the fermion momentum in the layer, and $c^*$ is the speed of light in the layer. Comparing \Eq{CCCA} and \Eq{NoMassCCCB}, I find
\be\label{NoMassWaveNumber}
qc^*=mc^2g\left(k/c\right)+V
\ee
and
\be
\alpha_q=-\dfrac{q}{q_x}\,.
\ee
\section{Results}
\label{Sec:RSE13}
\begin{figure}
\includegraphics*[width=\columnwidth]{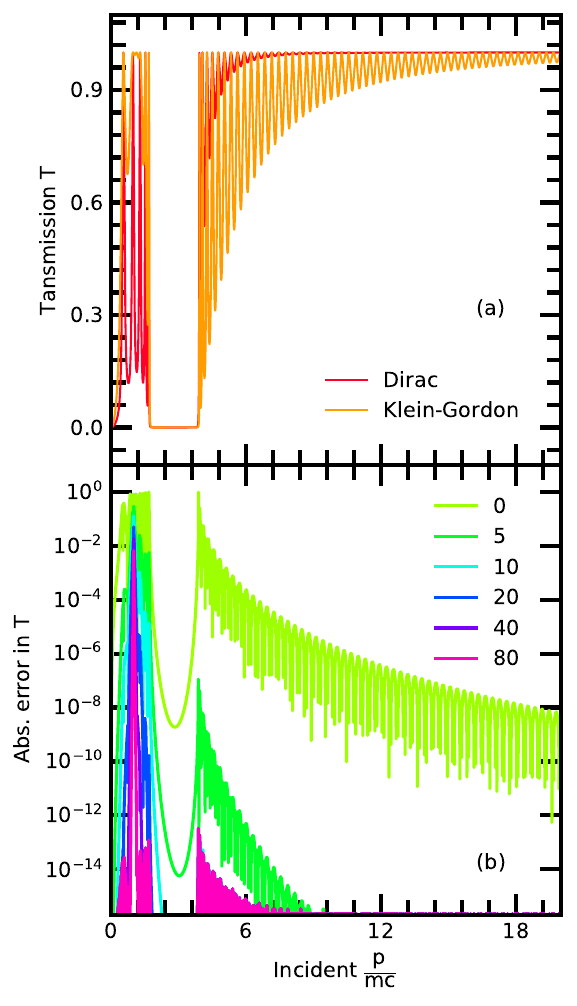}
\caption{(Color online) (a) Transmission through the potential barrier \Eq{CavityPotential} as a function of incident momentum $\mathnormal{p}$, for fermions (red) and bosons (orange). 
(b) Convergence of the fermion transmission with increasing number of relativistic correction Taylor expansion terms $0$, $5$, $10$, $20$, $40$, $80$ as labelled.}\label{fig:BarrierTunneling2a}
\end{figure}
\begin{figure}
\includegraphics*[width=\columnwidth]{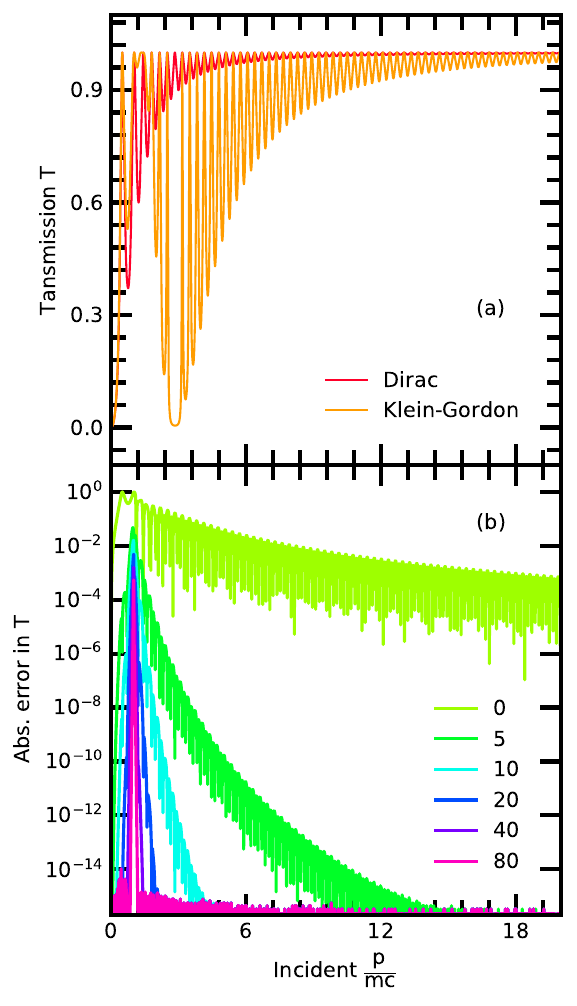}
\caption{(Color online) Repeat of \Fig{fig:BarrierTunneling2a}, except that now electron mass vanishes inside the barrier.}\label{fig:BarrierTunneling2bZeroMass}
\end{figure}
This section is restricted to quantum tunneling. For the structures I investigate in this section I introduce a length scale $a$ given in terms of electron mass $m$, the speed of light $c$, and $\hbar$
\be
a=5\dfrac{\hbar}{mc}\,.
\ee
In \App{App:A2} I investigate reflection of electrons from a sharp step in potential. In \App{App:A2} I find exact analytic agreement for $R$ between the FW and Dirac spinor representations. 

The striking feature of all my results is that I see exact analytic agreement between $T$ and $R$ calculated in the FW and Dirac spinor representations for all momentum $p$, even when $|p|>mc$. This finding is significant, as the Taylor expansion of $\mathcal{L}$ diverges when $|p|>mc$. I think that conservation of energy is restricting my evaluation of $\mathcal{\thickbar{L}}$ on $\widetilde{\Psi}^{(\pm)}$ to the particular values required to meet the physical boundary conditions for the Dirac Hamiltonian.

I model a resonant-tunneling diode to compare my WKB approximations of \Sec{Sec:RSE8} and \Sec{Sec:RSE8b} as the final example in this section.

For my numerical computations I use Python $3.8$.
\subsection{Tunneling through a barrier}
\label{SubSec:RSE12c}
The system under consideration is a potential barrier described by
\be
V\left(x\right)=-\left\{
\begin{array}{lc}  0 & \mbox{for}\ \  \left|x\right|>a\,,\\
3mc^2 & \mbox{for}\ \  \left|x\right|<a\,.\end{array}\right.
 \label{CavityPotential}
\ee
Since $V$ is negative, the momentum of the electron inside the barrier becomes imaginary when the following condition is met:
\be
-mc^2<E+V<+mc^2\,,
\ee
then the electron is quantum tunneling.

The analytic transmission for \Eq{CavityPotential} is derived in \App{App:A3} as:
\be\label{AnalyticRSE12c1}
T\left(k\right)=\dfrac{4\kappa^2}{4\kappa^2+\left(1-\kappa^2\right)^2\sin^2\left(\dfrac{2mqa}{\hbar}\right)}\,,
\ee
where I take $\kappa$ to be $\kappa_f$ in the FW representation, and $\kappa_d$ in the Dirac spinor representation. For the derivation of \App{App:A3} in the FW representation 
\be
\kappa^{\left(\pm\right)}_f=\dfrac{\alpha^{(\pm)}_k}{\alpha^{\left(\pm\right)}_q}=\dfrac{q_x}{k_x}\dfrac{E\pm mc^2}{E\pm mc^2+V}\,.
\ee
It was shown by the authors of Refs.~\cite{Calogeracos2010,Dosch1971} by deriving $T$ in the Dirac spinor representation of quantum tunneling
\be
\kappa^{\left(\pm\right)}_d=\dfrac{q_x}{k_x}\dfrac{E\pm mc^2}{E\pm mc^2+V}\,,
\ee
which is in exact analytic agreement with $\kappa^{\left(\pm\right)}_f$, as expected.

The numerical results demonstrating relativistic tunneling of electrons are shown in \Fig{fig:BarrierTunneling2a}. For comparison I also show the transmission of tunneling bosons with identical mass to the electron.

\Fig{fig:BarrierTunneling2a}(b) shows the absolute error calculated as the difference between $T$ evaluated using the Taylor expansions $g(k/c)$ in $\kappa$ or the Python $3.8$ $cmath$ module square root function.

The region of slow convergence around $|k/c|=1$ in \Fig{fig:BarrierTunneling2a}(b) is an artifact of the point in $g(k/c)$ about which I have taken my Taylor expansions.

\subsection{Tunneling through a barrier with vanishing electron mass inside the barrier}
\label{SubSec:RSE12d}
In this subsection I almost repeat the demonstration of \Sec{SubSec:RSE12c} except that now inside the barrier the electron mass is strictly vanishing.

Since $m=0$ when $|x|<a$, I must re-evaluate $\kappa_d$ for \Eq{AnalyticRSE12c1}:
\be
\kappa^{\left(\pm\right)}_d=\dfrac{q_x}{p_x}\dfrac{E\pm mc^2}{E+V}\,,
\ee
which is in exact analytic agreement with $\kappa^{\left(\pm\right)}_f$, as expected.

The numerical results demonstrating tunneling of relativistic electrons are shown in \Fig{fig:BarrierTunneling2bZeroMass}. For comparison I also show the transmission of tunneling bosons with identical mass to the electron.

In \Fig{fig:BarrierTunneling2bZeroMass}(b) the absolute error is calculated as the difference between $T$ evaluated using the Taylor expansions $g(k/c)$ in $\kappa$ or the Python $3.8$ $cmath$ module square root function.

The region of slow convergence around $|k/c|=1$ in \Fig{fig:BarrierTunneling2bZeroMass}(b) is an artifact of the point in $g(k/c)$ about which I have taken my Taylor expansions.
\subsection{Transmission through a resonant-tunneling diode using the WKB approximation}
\label{SubSec:RSE12e}
\begin{figure}
\includegraphics*[width=\columnwidth]{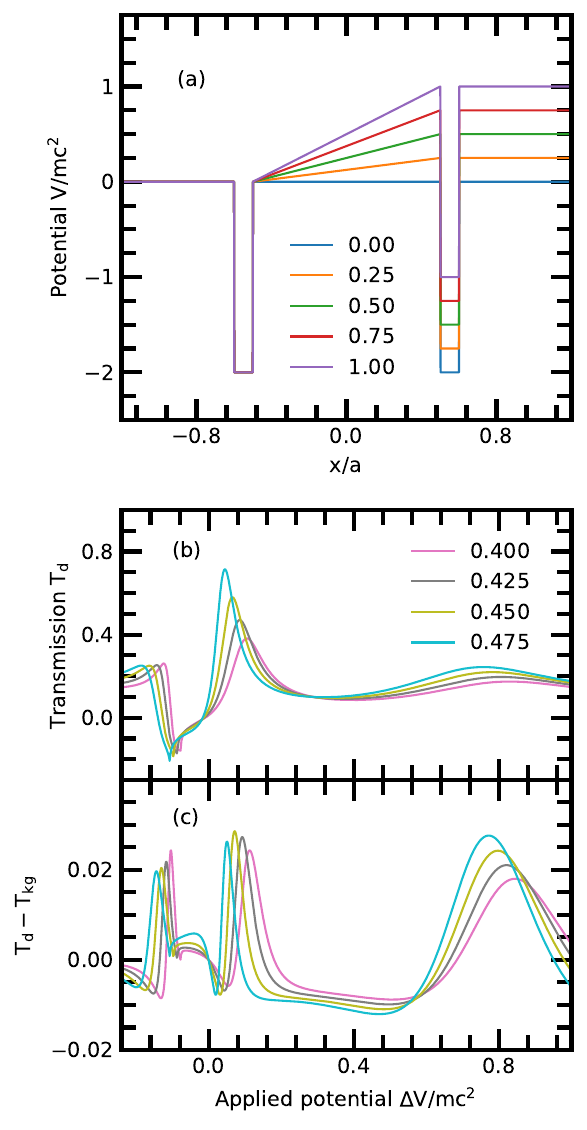}
\caption{(color online) (a) Potential profile for a representation of a resonant-tunneling diode under a range of bias potentials $\Delta V/mc^2$ 0.00, 0.25, 0.50, 0.75, 1.00 as labelled. (b) Transmission as a function of bias potential $\Delta V$ for a series of left hand side electron incident energies $E^{(i)}/mc^2$ 0.400, 0.425, 0.450, 0.475 as labelled, with $\Gamma_i$ taken to be $\pm\alpha^{\left(\pm\right)}_{i}$. (c) Difference in transmission, as a function of bias potential $\Delta V$, between using $\pm\alpha^{\left(\pm\right)}_{i}$ or $p_i$ for $\Gamma_i$.} \label{fig:Junction}
\end{figure}

I calculate the transmission through a resonant diode under a range of bias potentials $\Delta V$. I show a series of potentials describing increasing $\Delta V$ in \Fig{fig:Junction}(a). The transmission as a function of $\Delta V$ is given in \Fig{fig:Junction}(b) for a series of incident energies. I calculate transmission with the two different WKB approximations, derived in \Sec{Sec:RSE8} and \Sec{Sec:RSE8b}, for comparison in \Fig{fig:Junction}(c).

The resonant-tunneling diode that I consider includes a planar cavity with a smoothly varying potential. For the wavefunction to traverse the cavity the employment of WKB transfer matrices developed in \Sec{Sec:RSE8} and \Sec{Sec:RSE8b}\ are required:
\be
\begin{pmatrix}
    \dfrac{\exp{\left(- \dfrac{i}{\hbar}\int^{x_1}_{x_0} pdx\right)}}{\sqrt{\sfrac{\Gamma_0}{\Gamma_1}}}&0\\
    0&\dfrac{\exp{\left(+ \dfrac{i}{\hbar}\int^{x_1}_{x_0} pdx\right)}}{\sqrt{\sfrac{\Gamma_1}{\Gamma_0}}}
  \end{pmatrix}\begingroup\renewcommand*{\arraystretch}{3.0}\begin{pmatrix} \widetilde{a}_1 \\ \widetilde{b}_1 \end{pmatrix}\endgroup=\begingroup\renewcommand*{\arraystretch}{3.0}\begin{pmatrix} \widetilde{a}_0 \\ \widetilde{b}_0 \end{pmatrix}\endgroup\,,
\ee
$\Gamma_i$ is $p_i$ or $\pm\alpha^{\left(\pm\right)}_{i}$ depending on whether I use FW \Eq{OurFWEquation} or \Eq{FWEquation} inside the cavity. The subscript $0$ denotes evaluation at $x_0=-a/2$ the left hand side of the smoothly varying region for which the transfer matrix is applied, $1$ is at $x_1=+a/2$ the right hand side of the smoothly varying region. I define $\widetilde{a}_0$, $\widetilde{a}_1$, $\widetilde{b}_0$, and $\widetilde{b}_1$ in the same way as \Sec{Sec:RSE4}.

Since I have established in \Sec{SubSec:RSE12c} and \Sec{SubSec:RSE12d} that the Taylor series for $g(k/c)$ gives convergence to the square root function of the Python $3.8$ $cmath$ module, I dispense with the Taylor series and use the $cmath$ square root function to calculate the results in this subsection.

I see in \Fig{fig:Junction}(b) and (c) that as the energy of the incident electrons moves to higher values the first resonance of the diode moves towards lower values of $\Delta V$, as expected. I observe in \Fig{fig:Junction}(b) negative transmission as expected according to the Klein paradox \cite{Calogeracos2010,klein1929reflexion}.

In \Fig{fig:Junction}(b) and (c) I see no qualitative difference between my two WKB approximations, showing that exponential decay is the dominant factor in the cavity for this example.
\section{Conclusion}\label{sec:summary}
In this work I have developed Dirac fermion transfer matrix methods, for all energies, in the FW representations. I have also developed WKB approximations, to all orders, and discussed the validity of the approximations. I have introduced connection formulae between regions of real and imaginary fermion momentum.

I have demonstrated the applicability of my method to $2$D periodic structures for band-gap engineering \cite{Barbie2010,PhysRevB.77.115446,Arovas2010}.

I have alerted the reader to the limits of my methods by revealing a GF paradox that occurs when transforming between Dirac and FW representations. In spite of the paradox, I have rigorously derived the $1$D, $2$D and $3$D free space FW GFs. I note that since the FW \Eq{FWEquation} is a linear differential equation, the Dyson equations for the FW GFs are available for all types of Born approximation.

I have verified my methods and derivations for massive and effectively massless fermions of all momentum by analytic comparison with the Dirac spinor representation of fermion tunneling. I have shown that the FW representation is an exact description of Dirac fermions and is not restricted to semi-relativistic energies as first envisioned \cite{Foldy1950}.
\section{Acknowledgements}
This research was not funded by any research council grants. If you would like to support this research, please make a donation: \href{https://gofund.me/34294631}{https://gofund.me/34294631}. Contributions to support this research are greatly appreciated.

Future developments will involve the inclusion of RSE perturbation theory 
\cite{Doost2015,Doost2016,Muljarov2010,Doost2014,Armitage2014}. I plan to extend the derivation
of the normalization which I made in Ref.~\cite{Doost2014} to Dirac fermions described by the FW \Eq{FWEquation}. I acknowledge that as a sub-sequence to my rigorous normalization 
derivations in Ref.~\cite{Doost2014}, E. A. Muljarov
incorporated his zero frequency mode discovery into the generalized normalization. I acknowledge the first of the RSE waveguide articles Ref.~\cite{Armitage2014},
was predominantly the work of E. A. Muljarov.
\section{Declaration of competing interests}
I declare that I have no competing financial interests or personal relationships that could have influenced the work reported in this article. 
\appendix
\section{Discussion of commutation}
\label{App:A1}
In this appendix I will provide a discussion of
\be\label{LExpansion}
\mathcal{L}=-\mathcal{\thickbar{L}}c\hat{p}_x=-c\hat{p}_x\mathcal{\thickbar{L}}\,.
\ee
The operator $\mathcal{L}$ takes the form 
\be\
\mathcal{L}=mc^2\left[\dfrac{1}{2}\left[{\dfrac{\hat{p}}{mc}}\right]^2-\dfrac{1}{8}\left[{\dfrac{\hat{p}}{mc}}\right]^4+\dfrac{1}{16}\left[{\dfrac{\hat{p}}{mc}}\right]^6-...\right]\,,
\ee
where
\be
\dfrac{1}{2}\left[{\dfrac{\hat{p}}{mc}}\right]^2=-\dfrac{\hbar^2}{2m^2c^2}\left[\dfrac{d^2}{dx^2}+\dfrac{d^2}{dy^2}\right]\,,
\ee
\be
\dfrac{1}{8}\left[{\dfrac{\hat{p}}{mc}}\right]^4=\dfrac{\hbar^4}{8m^4c^4}\left[\dfrac{d^4}{dx^4}+2\dfrac{d^4}{dy^2dx^2}+\dfrac{d^4}{dy^4}\right]\,,
\ee
and the series of polynomial differential operators continues up to $n^{th}$ order. It is clear that I can write 
\be\label{EXTRACT1}
\dfrac{d^{n+m}}{dx^ndy^m}=\dfrac{d}{dx}\dfrac{d^{n+m-1}}{dx^{n-1}dy^m}=\dfrac{d^{n+m-1}}{dx^{n-1}dy^m}\dfrac{d}{dx}\,.
\ee
However, it is also the case that
\be\label{EXTRACT2}
\dfrac{d^n}{dy^n}f\left(x,y\right)=\dfrac{d}{dx}\dfrac{dx}{dy}\dfrac{d^{n-1}f}{dy^{n-1}}=\dfrac{d^{n-1}}{dy^{n-1}}\dfrac{dx}{dy}\dfrac{df}{dx}
\ee
when every term in $f$ is a function of not just $y$ but also $x$. I see from \Eq{EXTRACT1} and \Eq{EXTRACT2} that I can extract $d/dx$ from either side of the operator $\mathcal{L}$, in these cases.

When $f(x,y)=f(y)$, then in general
\be
\dfrac{d^n}{dy^n}f\left(y\right)\neq 0\,,
\ee
but
\be
\dfrac{d}{dx}f\left(y\right)=0\,,
\ee
therefore in general
\be\label{broken}
0\neq\dfrac{d^n}{dy^n}f\left(y\right)\neq\dfrac{d^{n-1}}{dy^{n-1}}\dfrac{dx}{dy}\dfrac{d}{dx} f\left(y\right)=0\,.
\ee
In the case described by \Eq{broken} I cannot extract $d/dx$ from either side of the operator $d^n/dy^n$.

\Eq{LExpansion} is true when I assume that $\mathcal{L}$ acts on a function of not just $y$ but also $x$.
\section{Reflection at a step potential in Foldy-Wouthuysen and Dirac spinor representations for comparison}
\label{App:A2}
Consider a fermion incident on a step at $x=0$. The wavefunction for $x<0$ is
\be\label{Step1}
\begin{pmatrix}E\pm mc^2\\ -cp_x\end{pmatrix}\exp{\left(+\dfrac{ip_xx}{\hbar}\right)}+
B\begin{pmatrix}E\pm mc^2\\ +cp_x\end{pmatrix}\exp{\left(-\dfrac{ip_xx}{\hbar}\right)}\,,
\ee
and for $x>0$
\be\label{Step2}
F\begin{pmatrix}E+V\pm mc^{*2}\\ -c^*q_x\end{pmatrix}\exp{\left(+\dfrac{iq_xx}{\hbar}\right)}\,.
\ee
Using wavefunction continuity at $x=0$, I find:
\be\label{Step3}
\big(1-B\big)cp_x=Fc^*q_x\,,
\ee
\be\label{Step4}
\big(1+B\big)\big(E\pm mc^2\big)=F\big(E+V\pm mc^{*2}\big)\,.
\ee
Dividing \Eq{Step3} by \Eq{Step4} gives
\be
\dfrac{1-B}{1+B}=\dfrac{q_xc^*}{p_xc}\dfrac{E\pm mc^2}{E+V\pm mc^{*2}}=\kappa^{\left(\pm\right)}_d\,.
\ee

Now I turn to the FW representation. The corresponding wavefunction for $x<0$ is
\be
\exp{\left(+\dfrac{ip_xx}{\hbar}\right)}+B\exp{\left(-\dfrac{ip_xx}{\hbar}\right)}\,,
\ee
and for $x>0$
\be
F\exp{\left(+\dfrac{iq_xx}{\hbar}\right)}\,.
\ee
Applying my boundary condition at $x=0$, continuity of wavefunction and continuity of wavefunction with respect to $\mathcal{\thickbar{L}}$, I obtain
\be\label{StepFW1}
\big(1-B\big)\alpha^{\left(\pm\right)}_0=F\alpha^{\left(\pm\right)}_1\,,
\ee
\be\label{StepFW2}
1+B=F\,.
\ee
Dividing \Eq{StepFW1} by \Eq{StepFW2} gives
\be
\dfrac{1-B}{1+B}=\dfrac{\alpha^{\left(\pm\right)}_1}{\alpha^{\left(\pm\right)}_0}=\dfrac{p_xc}{q_xc^*}\dfrac{E+V\pm mc^{*2}}{E\pm mc^2}=\kappa^{\left(\pm\right)}_f\,.
\ee

In \Sec{Sec:RSE2} I derived $R=|B|^2$, therefore
\be\label{StepRVerification}
R=\left|\dfrac{1-\kappa}{1+\kappa}\right|^2\,,
\ee
where I use $\kappa^{(\pm)}_d$ for the Dirac equation and $\kappa^{\left(\pm\right)}_f$ for the FW equation. Substituting either $\kappa^{\left(\pm\right)}_f$ or $\kappa^{\left(\pm\right)}_d$ into \Eq{StepRVerification} gives analytically identical $T$ and $R$ between the Dirac and FW representation, as expected. I also calculated $\kappa^{\left(\pm\right)}_d$ in the case of $m$ vanishing for $x>0$ and found exact analytic agreement in $T$ and $R$ between the Dirac and FW representations, as expected.

I calculate, for useful comparison, the transmission of bosons with the following Klein-Gordon result
\be
\kappa_{kg}=\dfrac{q_xc^*}{p_xc}\,.
\ee
\section{Examination of a massless fermion tunneling through a magnetic delta barrier in the Foldy-Wouthuysen representation}
\label{App:A3a}
Consider a massless fermion tunneling through a magnetic delta barrier $B=B_0\delta\left(x\right)\hat{z}$, where $B=\nabla\times A$. The barrier is described by $A=0$ except for $A=A_y\hat{y}\neq 0$ when $x>0$. The Dirac spinor wavefunction is a  solution of
\be\label{DIRACmag}
\left[c{\bf\alpha}\cdot\left(\hat{p}-A\right)\right]\Psi=E\Psi\,.
\ee
Using the approach of \Sec{Sec:RSE1}, \Eq{DIRACmag} can be transformed to
\be\label{DIRACmagDiag}
\left[\mp c\sqrt{\left(\hat{p}-A\right)^2}\right]\widetilde{\Psi}^{\left(\pm\right)}=E\widetilde{\Psi}^{\left(\pm\right)}\,.
\ee
Taylor expanding \Eq{DIRACmagDiag} around $\hat{p}=0$ gives
\be\label{DIRACmagLinear}
\left[\mp \mathcal{M} - \left(E\pm cA_y\right)\right]\widetilde{\Psi}^{\left(\pm\right)}=0\,,
\ee
where $\mathcal{M}$ is the corresponding linear differential operator. Using the approach of \Sec{Sec:RSE3} I calculate, that at the barrier, the wavefunction is continuous and also continuous with respect to $\mathcal{\thickbar{M}}$, where I define
\be
\mathcal{\thickbar{M}}\widetilde{\Psi}^{\left(\pm\right)}=\pm\dfrac{E\pm cA_y}{cp_x}\widetilde{\Psi}^{\left(\pm\right)}\,.
\ee

I now return to the Dirac spinor representation. In the region $A=A_y\hat{y}\neq 0$, the general solution for $\hat{p}=\pm q_x\hat{x}$ is given by \cite{masir2008direction, masir2008wavevector}
\be\label{PsiMag}
\Psi=\begin{pmatrix}+E \\ \pm cq_x-icA_y\end{pmatrix}\exp{\left(\pm\dfrac{iq_xx}{\hbar}\right)}\,,
\ee
where $cq_x=\sqrt{E^2-c^2A_y^2}$. The reflection is calculated by matching wavefunctions at the step in vector potential, similar to \App{App:A3a} for a step in electrostatic potential.

Making use of my boundary conditions I find the reflection coefficient $R$ takes the form of \Eq{StepRVerification}, with
\be\label{kappadA}
\kappa_d=\dfrac{c^*}{c}\dfrac{q_x-iA_y}{p_x}
\ee
for the Dirac spinor representation and
\be\label{kappafA}
\kappa_f=\dfrac{cp_x}{c^*q_x}\dfrac{E\pm c^*A_y}{E}
\ee
for the FW representation. $\hat{p}=p_x\hat{x}=(E/c)\hat{x}$ when $A=0$.  

The conflict between \Eq{kappadA} and \Eq{kappafA} indicates $R$ calculated in the FW representation does not reproduce $R$ calculated in the Dirac spinor representation. This suggests that FW wavefunctions are unsuitable for calculating tunneling through magnetic barriers. However, this conflict might be resolved by employing the form of the FW transformation in Ref.~\cite{Foldy1950} that derives the Pauli equation from the Dirac equation in the non-relativistic limit.

\section{Analytic Green's functions in one dimension}
\label{App:A3}
Consider
\be
\left[\mp\mathcal{L}-\left(V+E\pm mc^2\right)\right]G^{\left(\pm\right)}\left(x,x'\right)=\delta\left(x-x'\right)\,.\label{A1}
\ee
Let $\widetilde{\Psi}_L$ and $\widetilde{\Psi}_R$ be the solutions of \Eq{FWEquation} which separately satisfy the boundary conditions
at $x=-a$ and $x=+a$. The GF of \Eq{A1} for the case of a $1$D barrier
can be written as
\be
G^{\left(\pm\right)}\left(x,x'\right)=\dfrac{\widetilde{\Psi}^{\left(\pm\right)}_L\left(x_<\right)\widetilde{\Psi}^{\left(\pm\right)}_R\left(x_>\right)}{{W}^{\left(\pm\right)}}\,,\label{A2}
\ee
where $x_<=\min{(x,x')}$ and $x_>=\max{(x,x')}$ and
\be
{W}^{\left(\pm\right)}=\mp \widetilde{\Psi}^{\left(\pm\right)}_Lic\hbar\mathcal{\thickbar{L}}\widetilde{\Psi}^{\left(\pm\right)}_R\pm\widetilde{\Psi}^{\left(\pm\right)}_Ric\hbar\mathcal{\thickbar{L}}\widetilde{\Psi}^{\left(\pm\right)}_L\label{A3}
\ee
is the Wronskian, which does not depend on $x$.

To prove \Eq{A2} and \Eq{A3}, note that by \Eq{A1}
\be
\mp\lim_{\epsilon\rightarrow 0}\int^{x'+\epsilon}_{x'-\epsilon} \mathcal{L}G^{\left(\pm\right)}\left(x,x'\right)dx=1\,,\label{A4}
\ee
which implies
\be
\mp ic\hbar\mathcal{\thickbar{L}}G^{\left(\pm\right)}\left(x_+,x'\right)\pm ic\hbar\mathcal{\thickbar{L}}G^{\left(\pm\right)}\left(x_-,x'\right)=1\,,\label{A5}
\ee
then, applying $\mathcal{\thickbar{L}}$ to \Eq{A2} written out explicitly
\be
\mathcal{\thickbar{L}}G\left(x,x'\right)=\dfrac{1}{{W}}\left\{
\begin{array}{lc}\mathcal{\thickbar{L}}\widetilde{\Psi}_L\big(x\big)\widetilde{\Psi}_R\big(x'\big)  & \mbox{for}\ \  x<x'\,,\\
\widetilde{\Psi}_L\big(x'\big)\mathcal{\thickbar{L}}\widetilde{\Psi}_R\big(x\big) & \mbox{for}\ \  x>x'\,,\end{array}\right.
\ee
and substituting into \Eq{A5} gives \Eq{A3}.

I now calculate the GF for a rectangular barrier of width $2a$.
\be \widetilde{\Psi}_1=\left\{
\begin{array}{lll}
\exp{\bigg(+\dfrac{imkx}{\hbar}\bigg)}&x>+a\,,\\
C\exp{\left(+\dfrac{im^*qx}{\hbar}\right)}+D\exp{\left(-\dfrac{im^*qx}{\hbar}\right)}&\left|x\right|<a\,,\\
A\exp{\bigg(+\dfrac{imkx}{\hbar}\bigg)}+B\exp{\bigg(-\dfrac{imkx}{\hbar}\bigg)}&x<-a\,.
\end{array} \right.
\ee
Continuity of $\widetilde{\Psi}_1(x)$ and  continuity of $\widetilde{\Psi}_1(x)$ with respect to $\mathcal{\thickbar{L}}$ gives
\be
C=\dfrac{\alpha_q+\alpha_k}{2\alpha_q}\exp{\bigg(+\dfrac{imka}{\hbar}\bigg)}\exp{\bigg(-\dfrac{im^*qa}{\hbar}\bigg)}\,,
\ee
\be
D=\dfrac{\alpha_q-\alpha_k}{2\alpha_q}\exp{\bigg(+\dfrac{imka}{\hbar}\bigg)}\exp{\bigg(+\dfrac{im^*qa}{\hbar}\bigg)}\,.
\ee
$\widetilde{\Psi}_2$ the left hand solution is given by $\widetilde{\Psi}_2(+x)=\widetilde{\Psi}_1(-x)$. Therefore I evaluate \Eq{A2} to give $G(+a,-a)$:
\be\label{G_pa_na}
\dfrac{i\alpha_q}{ic\hbar\left(\alpha^2_q+\alpha^2_k\right)\sin{\left(\dfrac{2m^*qa}{\hbar}\right)}-2c\hbar\alpha_q\alpha_k\cos{\left(\dfrac{2m^*qa}{\hbar}\right)}}\,.
\ee
In \Eq {G_pa_na}, the Wronskian is given by 
\be
{W}=2ic\hbar\alpha_q\left(C^2-D^2\right)\,.
\ee

The transmission for a barrier embedded in a vacuum is given as the ratio of $G(+a,-a)$ and the free space GF $G^{fs}(+a,-a)$:
\be
T(k)=|2c\hbar\alpha_k G\left(+a,-a\right)|^2=\left|\dfrac{G\left(+a,-a\right)}{G^{fs}\left(+a,-a\right)}\right|^2\,.
\ee
The free space GF can be written as 
\be
\left[\mp\mathcal{L}-\left(E\pm mc^2\right)\right]G^{(\pm)}\left(x,x'\right)=\delta\left(x-x'\right)\,,\label{B1}
\ee
therefore by symmetry, when $x'=0$,
\be\label{FWGreensFunction1D}
G^{(\pm)}(x,0)=A^{\left(\pm\right)}\left\{
\begin{array}{lc}\exp{\left(-\dfrac{imkx}{\hbar}\right)}  & \mbox{for}\ \  x<0\,,\\
\exp{\left(+\dfrac{imkx}{\hbar}\right)} & \mbox{for}\ \  x>0\,.\end{array}\right.
\ee
Integrating \Eq{B1} across $\left[0_{-},0_{+}\right]$ gives
\be
\mp{i}{c\hbar A^{\left(\pm\right)}}\mathcal{\thickbar{L}}\exp{\left(+\dfrac{imk0_+}{\hbar}\right)}\pm{i}{c\hbar A^{\left(\pm\right)}}\mathcal{\thickbar{L}}\exp{\left(-\dfrac{imk0_-}{\hbar}\right)}=1
\ee
and letting $0_{+}\rightarrow 0$ and $0_{-}\rightarrow 0$, I find 
\be\label{FWGreensFunction1DAFactor}
A^{\left(\pm\right)}=\dfrac{1}{2ic\hbar\alpha^{\left(\pm\right)}_k}\,.
\ee
\section{Green's function paradox}
\label{App:A4}
I reveal a GF paradox occurring when transforming between the FW and Dirac representations.

Let
\be
\left[H-\left(E+V\right)\right]\mathcal{G}\left(\br,\br'\right)=\delta\left(\br-\br'\right)\hat{1}\,,
\ee
therefore
\begin{multline}
e^{+iS}\left[H-\left(E+V\right)\right]e^{-iS}e^{+iS}\left[\mathcal{G}\left(\br,\br'\right)\right]e^{-iS}
\\
=e^{+iS}\hat{1}e^{-iS}\delta\left(\br-\br'\right)\,.
\end{multline}
However since
\be
e^{+iS}\hat{1}e^{-iS}\delta\left(\br-\br'\right)=\hat{1}\delta\left(\br-\br'\right)\,,
\ee
and since I have already seen in \Sec{Sec:RSE1}
\be
e^{+iS}\left[H-\left(E+V\right)\right]e^{-iS}=\left[\mp\mathcal{L}-\left(V+E\pm mc^2\right)\right]\hat{1}\,,
\ee
I have
\begin{multline}\label{DiagonalGF}
e^{+iS}\left[\mathcal{G}\left(\br,\br'\right)\right]e^{-iS}=G^{\left(\pm\right)}\left(\br,\br'\right)\hat{1}
\\
\implies e^{-iS}G^{\left(\pm\right)}\left(\br,\br'\right)e^{+iS}=\mathcal{G}\left(\br,\br'\right)\hat{1}\,.
\end{multline}
\Eqs{DiagonalGF} imply that the valid GFs of the Dirac equation are diagonal. Now consider the identity
\be\label{BI}
\left[H-\left(E+V\right)\right]\left[H+\left(E+V\right)\right]K\left(\br,\br'\right)=\hat{1}\delta\left(\br-\br'\right)\,,
\ee
so that
\be\label{BI2} 
\left[H+\left(E+V\right)\right]K\left(\br,\br'\right)=\mathcal{G}\left(\br,\br'\right)\,.
\ee
Multiplying out the brackets in \Eq{BI} I obtain
\be\label{FreeSpaceKleinGordon}
-c^2\hbar^2\left[\nabla^2+\left(\dfrac{p}{\hbar}\right)^2\right]K\left(\br,\br'\right)=\delta\left(\br-\br'\right)\,.
\ee
I see from \Eq{FreeSpaceKleinGordon} that in $1$D free space ($V=0$):
\be\label{KG1DFS}
c^2\hbar^2K\left(x,x'\right)=-\dfrac{\hbar\exp{\left(\dfrac{ip\left|x-x'\right|}{\hbar}\right)}}{2ip}\,,
\ee
in $2$D free space:
\be\label{KG2DFS}
c^2\hbar^2K\left(\rho,\rho'\right)=\dfrac{i}{4}H^{(1)}_0\left(\dfrac{p\left|\rho-\rho'\right|}{\hbar}\right)\,,
\ee
and in $3$D free space:
\be\label{KG3DFS}
c^2\hbar^2K\left(\br,\br'\right)=\dfrac{\exp{\left(\dfrac{ip\left|\br-\br'\right|}{\hbar}\right)}}{4\pi\left|\br-\br'\right|}\,.
\ee
Substituting the identity \Eq{BI2} into \Eq{DiagonalGF}
\be
e^{+iS}\left[H+\left(E+V\right)\right]e^{-iS}K\left(\br,\br'\right)=G^{(\pm)}\left(\br,\br'\right)\hat{1}
\ee
and so 
\be\label{Eig2}
\left[\sqrt{\hat{p}^2c^2+m^2c^4}+\left(E+V\right)\right]K\left(\br,\br'\right)=G^{\left(\pm\right)}\left(\br,\br'\right)\hat{1}\,.
\ee
By inspection of \Eq{Eig2} and the energy momentum relation $(E+V)=\sqrt{p^2c^2+m^2c^4}$,
\be\label{Gpm2EV}
G^{\left(\pm\right)}\left(\br,\br'\right)=2\left(E+V\right)K\left(\br,\br'\right)
\ee
and substituting \Eq{Gpm2EV} into \Eq{DiagonalGF} I also have 
\be\label{KEY}
\mathcal{G}\left(\br,\br'\right)=2\left(E+V\right)K\left(\br,\br'\right)\,.
\ee

Let us test \Eq{Gpm2EV} and \Eq{KEY} by making an analytic comparison between the GF for $1$D free space \Eq{FWGreensFunction1D} and the GF composed of \Eq{Gpm2EV} and \Eq{KG1DFS}:
\be
2EK=-E\dfrac{\exp{\left(\dfrac{ip\left|x-x'\right|}{\hbar}\right)}}{i\hbar pc^2}\neq \dfrac{\exp{\left(\dfrac{ip\left|x-x'\right|}{\hbar}\right)}}{2ic\hbar\alpha_k}
\ee
and so I have an apparent paradox.

Let us attempt to resolve the paradox. I have shown in \App{App:A3} that there are two possible forms of the FW GF:
\be
G^{\left(\pm\right)}\left(x,x'\right)=\dfrac{\exp{\left(\dfrac{ip\left|x-x'\right|}{\hbar}\right)}}{2ic\hbar\alpha^{\left(\pm\right)}_k}
\ee
for an electron delta source ($-$) or a positron source ($+$). If I add my two expressions $G^{(\pm)}(x,x')$ I find
\be
G^{\left(-\right)}\left(x,x'\right)+G^{\left(+\right)}\left(x,x'\right)=2EK\left(x,x'\right)\,.
\ee
However if I try to prove that in general
\be
\mathcal{G}\left(\br,\br'\right)=G^{\left(-\right)}\left(\br,\br'\right)+G^{\left(+\right)}\left(\br,\br'\right)\,,
\ee
I arrive at a contradiction. The Klein-Gordon GF has the continuity conditions of the Klein-Gordon equation, $(E+V)$ can be a discontinuous function, therefore $2(E+V)K$ can be discontinuous at boundaries. Therefore \Eq{KEY} implies $\mathcal{G}$ may be discontinuous. $\mathcal{G}$ has the same boundary conditions as $\Psi$, \Eq{KEY} may violate the boundary conditions for $\mathcal{G}$.

Let us now discuss a possible cause of the paradox. To evaluate the FW transformations defined to be 
\be
e^{\pm iS}=\cos\theta\pm{\bf\beta}{\bf\alpha}\cdot\dfrac{\hat{p}}{|\hat{p}|}\sin\theta\,,
\ee
I am required to know the momentum $\hat{p}$ also at the delta source. To make this calculation of $\hat{p}$ at the delta source consider \Eq{FWGreensFunction1D} from \App{App:A3},
\be\label{FWGreensFunction1D_App}
G\left(x,0\right)=\dfrac{1}{2ic\hbar\alpha_k}\left\{
\begin{array}{lc}  \exp{\left(-\dfrac{imkx}{\hbar}\right)}  & \mbox{for}\ \  x<0\,,\\
\exp{\left(+\dfrac{imkx}{\hbar}\right)} & \mbox{for}\ \  x>0\,.\end{array}\right.
\ee
I see from \Eq{FWGreensFunction1D_App} that
\be\label{FWGreensFunction1D_App1}
p_x\left(x=0\right)G\left(x,0\right)=\lim_{0_{\pm}\rightarrow 0}\hat{p}G\left(0_{\pm},0\right)={\pm}mkG\left(0,0\right)\,.
\ee
\Eq{FWGreensFunction1D_App1} indicates that when $x=x'$, the momentum $\hat{p}$ is undefined. Therefore at the delta source the FW transformation is undefined. For $1$D space it might be impossible to evaluate $e^{\pm iS}$ at the delta source.
\section{Correct free space Foldy-Wouthuysen Green's functions in two and three dimensions}
\label{App:A5}
I begin with the $3$D FW GF in free space 
\be\label{SAB0} G^{(\pm)}_{\rm 3D}\left(\br, \br'\right) = -\dfrac{p}{c\hbar^2\alpha^{(\pm)}_k}\dfrac{1}{4\pi\left|\br - \br'\right|} e^{ip|\br - \br'|/\hbar}\,,\ee
which at the end of this \App{App:A5}, I will show to be consistent with the proven $1$D FW GF, \Eq{FWGreensFunction1D} and \Eq{FWGreensFunction1DAFactor}. \Eq{SAB0} is the solution of
\be
\left[\mp\mathcal{L}-\left(E\pm mc^2\right)\right]G^{(\pm)}_{\rm 3D}\left(\br, \br'\right)=\delta_{\rm 3D}\left(\br-\br'\right)\,.
\ee
\Eq{SAB0} can be thought of as an outgoing exponential plane FW wavefunction spreading out over a sphere centered on the source, as we move further from the source. In \Eq{SAB0} I make a choice of amplitude to obtain agreement with the correct $1$D FW GF. 

To derive the $1$D free space FW GF, I integrate \Eq{SAB0} in the plane $(y,z)$ and find
\be\label{SAB3} G^{(\pm)}_{\rm 1D}\left(x, x'\right)=\int^{+\infty}_{-\infty}\int^{+\infty}_{-\infty}G^{(\pm)}_{\rm 3D}\left(\br, \br'\right)dydz\,,\ee
where $r=\left(x,y,z\right)$. \Eq{SAB3} is a consequence of
\be\label{DeltaSAB3} \delta_{\rm 1D}\left(x-x'\right)=\int^{+\infty}_{-\infty}\int^{+\infty}_{-\infty}\delta_{\rm 3D}\left(\br-\br'\right)dydz\,.\ee
Let $\rho=\sqrt{y^2+z^2}$ in the cylindrical coordinates $(\rho,\theta,x)$, and write $\xi=\sqrt{(x-x')^2+\rho^2}$, so that \Eq{SAB3} becomes
\be\label{SAB4}
 G^{(\pm)}_{\rm 1D}\left(x, x'\right)=-\dfrac{p}{c\hbar^2\alpha^{(\pm)}_k}\int^{2\pi}_{0}\int^{\infty}_{0}\dfrac{1}{4\pi\xi} e^{ip\xi/\hbar}\rho d\rho d\theta\,.
\ee
Integrating \Eq{SAB4} with respect to $\theta$ and changing variable to $\xi$ gives, using the chain rule 
\be
d\xi=\dfrac{\rho}{\sqrt{(x-x')^2+\rho^2}}d\rho=\dfrac{\rho}{\xi}d\rho\,,
\ee
noting in the integration limits that I have $\xi=|x-x'|$ as a minimum when $\rho=0$:
\be\label{SAB5}
 G^{(\pm)}_{\rm 1D}\left(x, x'\right)=-\dfrac{p}{c\hbar^2\alpha^{(\pm)}_k}\int^{\infty}_{|x-x'|}\dfrac{1}{2} e^{ip\xi/\hbar}d\xi\,.
\ee
\Eq{SAB5} can be integrated while dropping the oscillating upper integration limit at $\xi=\infty$, assuming the contributions at $\xi=\infty$ cancel out leading to a zero net contribution,
\be\label{IntegralBad}
\int^{\infty}_{|x-x'|}\dfrac{1}{2}e^{ip\xi/\hbar}d\xi=\left[\dfrac{\hbar}{2ip}e^{ip\xi/\hbar}\right]^{\infty}_{|x-x'|}=-\dfrac{\hbar}{2ip}e^{ip|x-x'|/\hbar}\,.
\ee
Hence I arrive at the required result, \Eq{FWGreensFunction1D} combined with \Eq{FWGreensFunction1DAFactor}:
\be\label{SAB6}
G^{(\pm)}_{\rm 1D}\left(x, x'\right)=\dfrac{p}{c\hbar^2\alpha^{(\pm)}_k}\dfrac{\hbar}{2ip}e^{ip|x-x'|/\hbar}=\dfrac{e^{ip|x-x'|/\hbar}}{2ic\hbar\alpha^{(\pm)}_k}\,.
\ee

The integration in \Eq{IntegralBad} can be seen clearly when I take $p$ to be complex with even only a vanishingly small imaginary part so that the upper integration limit is an exponentially decaying function, rather than just an oscillating function.

An interesting comparison can be made between the FW GF \Eqs{SAB0} and \Eqs{SAB6} and their Klein-Gordon counterparts, \Eqs{KG3DFS} and \Eqs{KG1DFS}, which were derived in \App{App:A4}. The derivation in this \App{App:A5} can be repeated, as a simple checking exercise, for \Eqs{KG3DFS} and \Eqs{KG1DFS}. These comparisons and derivations, repeated in $1$D and $2$D space, suggest that the $2$D confined free space FW GF is given by
\be\label{CSAB0} G^{(\pm)}_{\rm 2D}\left(\br, \br'\right) = -\dfrac{p}{c\hbar^2\alpha^{(\pm)}_k}\dfrac{i}{4}H^{(1)}_0\left(p\left|\br - \br'\right|/\hbar\right)\,.
\ee
\Eq{SAB0} and \Eq{CSAB0} can be simplified with
\be
\dfrac{p}{c\hbar^2\alpha^{(\pm)}_k}=-\dfrac{E\mp mc^2}{c^2\hbar^2}\,.
\ee

I note that since the FW \Eq{FWEquation} is a linear differential equation, the Dyson equations for the FW GFs are available for all types of Born approximation \cite{Doost2015,Doost2016}. There might be cases where the decoupling of electron and positron states by the FW approach could simplify or eliminate certain divergences in the Dyson series, potentially addressing aspects of the renormalization problem.
\bibliographystyle{elsarticle-num-names}
\bibliography{cas-refs}

\begin{thebibliography}{23}
\expandafter\ifx\csname natexlab\endcsname\relax\def\natexlab#1{#1}\fi
\providecommand{\url}[1]{\texttt{#1}}
\providecommand{\href}[2]{#2}
\providecommand{\path}[1]{#1}
\providecommand{\DOIprefix}{doi:}
\providecommand{\ArXivprefix}{arXiv:}
\providecommand{\URLprefix}{URL: }
\providecommand{\Pubmedprefix}{pmid:}
\providecommand{\doi}[1]{\href{http://dx.doi.org/#1}{\path{#1}}}
\providecommand{\Pubmed}[1]{\href{pmid:#1}{\path{#1}}}
\providecommand{\bibinfo}[2]{#2}
\ifx\xfnm\relax \def\xfnm[#1]{\unskip,\space#1}\fi
\bibitem[{Barbier et~al.(2010)Barbier, Vasilopoulos, and Peeters}]{Barbie2010}
\bibinfo{author}{M.~Barbier}, \bibinfo{author}{P.~Vasilopoulos}, \bibinfo{author}{F.~Peeters},
\newblock \bibinfo{title}{Extra dirac points in the energy spectrum for superlattices on single-layer graphene},
\newblock \bibinfo{journal}{Physical Review B} \bibinfo{volume}{81} (\bibinfo{year}{2010}) \bibinfo{pages}{075438}. \DOIprefix\doi{https://doi.org/10.1103/PhysRevB.81.075438}.
\bibitem[{Barbier et~al.(2008)Barbier, Peeters, Vasilopoulos, and Pereira}]{PhysRevB.77.115446}
\bibinfo{author}{M.~Barbier}, \bibinfo{author}{F.~Peeters}, \bibinfo{author}{P.~Vasilopoulos}, \bibinfo{author}{J.~M. Pereira},
\newblock \bibinfo{title}{Dirac and klein-gordon particles in one-dimensional periodic potentials},
\newblock \bibinfo{journal}{Phys. Rev. B} \bibinfo{volume}{77} (\bibinfo{year}{2008}) \bibinfo{pages}{115446}. \DOIprefix\doi{https://doi.org/10.1103/PhysRevB.77.115446}.
\bibitem[{Arovas et~al.(2010)Arovas, Brey, Fertig, Kim, and Ziegler}]{Arovas2010}
\bibinfo{author}{D.~Arovas}, \bibinfo{author}{L.~Brey}, \bibinfo{author}{H.~Fertig}, \bibinfo{author}{E.-A. Kim}, \bibinfo{author}{K.~Ziegler},
\newblock \bibinfo{title}{Dirac spectrum in piecewise constant one-dimensional (1d) potentials},
\newblock \bibinfo{journal}{New Journal of Physics} \bibinfo{volume}{12} (\bibinfo{year}{2010}) \bibinfo{pages}{123020}. \DOIprefix\doi{https://doi.org/10.1088/1367-2630/12/12/123020}.
\bibitem[{Guinea et~al.(2010)Guinea, Katsnelson, and Geim}]{Guinea2010}
\bibinfo{author}{F.~Guinea}, \bibinfo{author}{M.~Katsnelson}, \bibinfo{author}{A.~Geim},
\newblock \bibinfo{title}{Energy gaps and a zero-field quantum hall effect in graphene by strain engineering},
\newblock \bibinfo{journal}{Nature Physics} \bibinfo{volume}{6} (\bibinfo{year}{2010}) \bibinfo{pages}{30--33}. \DOIprefix\doi{https://doi.org/10.1038/nphys1420}.
\bibitem[{Novoselov et~al.(2016)Novoselov, Mishchenko, Carvalho, and Neto}]{Novoselov2016}
\bibinfo{author}{K.~S. Novoselov}, \bibinfo{author}{A.~Mishchenko}, \bibinfo{author}{A.~Carvalho}, \bibinfo{author}{A.~H.~C. Neto},
\newblock \bibinfo{title}{2d materials and van der waals heterostructures},
\newblock \bibinfo{journal}{Science} \bibinfo{volume}{353} (\bibinfo{year}{2016}). \DOIprefix\doi{https://doi.org/10.1126/science.aac9439}.
\bibitem[{Wan et~al.(2015)Wan, Gu, Dang, Itoh, Wang, Sasaki, Kondo, Koga, Yabuki, Snyder et~al.}]{Wan2015}
\bibinfo{author}{C.~Wan}, \bibinfo{author}{X.~Gu}, \bibinfo{author}{F.~Dang}, \bibinfo{author}{T.~Itoh}, \bibinfo{author}{Y.~Wang}, \bibinfo{author}{H.~Sasaki}, \bibinfo{author}{M.~Kondo}, \bibinfo{author}{K.~Koga}, \bibinfo{author}{K.~Yabuki}, \bibinfo{author}{G.~J. Snyder}, et~al.,
\newblock \bibinfo{title}{Flexible n-type thermoelectric materials by organic intercalation of layered transition metal dichalcogenide tis 2},
\newblock \bibinfo{journal}{Nature materials} \bibinfo{volume}{14} (\bibinfo{year}{2015}) \bibinfo{pages}{622--627}. \DOIprefix\doi{https://doi.org/10.1038/nmat4251}.
\bibitem[{Song et~al.(2013)Song, Wu, and Guo}]{Song2013}
\bibinfo{author}{Y.~Song}, \bibinfo{author}{H.-C. Wu}, \bibinfo{author}{Y.~Guo},
\newblock \bibinfo{title}{Negative differential resistances in graphene double barrier resonant tunneling diodes},
\newblock \bibinfo{journal}{Applied Physics Letters} \bibinfo{volume}{102} (\bibinfo{year}{2013}) \bibinfo{pages}{093118}. \DOIprefix\doi{https://doi.org/10.1063/1.4794952}.
\bibitem[{Sonin(2009)}]{Sonin2009}
\bibinfo{author}{E.~Sonin},
\newblock \bibinfo{title}{Effect of klein tunneling on conductance and shot noise in ballistic graphene},
\newblock \bibinfo{journal}{Physical Review B} \bibinfo{volume}{79} (\bibinfo{year}{2009}) \bibinfo{pages}{195438}. \DOIprefix\doi{https://doi.org/10.1103/PhysRevB.79.195438}.
\bibitem[{Tudorovskiy et~al.(2012)Tudorovskiy, Reijnders, and Katsnelson}]{Tudorovskiy2012}
\bibinfo{author}{T.~Tudorovskiy}, \bibinfo{author}{K.~Reijnders}, \bibinfo{author}{M.~I. Katsnelson},
\newblock \bibinfo{title}{Chiral tunneling in single-layer and bilayer graphene},
\newblock \bibinfo{journal}{Physica Scripta} \bibinfo{volume}{2012} (\bibinfo{year}{2012}) \bibinfo{pages}{014010}. \DOIprefix\doi{https://doi.org/10.1088/0031-8949/2012/T146/014010}.
\bibitem[{Zalipaev et~al.(2015)Zalipaev, Linton, Croitoru, and Vagov}]{Zalipaev2015}
\bibinfo{author}{V.~Zalipaev}, \bibinfo{author}{C.~Linton}, \bibinfo{author}{M.~Croitoru}, \bibinfo{author}{A.~Vagov},
\newblock \bibinfo{title}{Resonant tunneling and localized states in a graphene monolayer with a mass gap},
\newblock \bibinfo{journal}{Physical Review B} \bibinfo{volume}{91} (\bibinfo{year}{2015}) \bibinfo{pages}{085405}. \DOIprefix\doi{https://doi.org/10.1103/PhysRevB.91.085405}.
\bibitem[{Foldy and Wouthuysen(1950)}]{Foldy1950}
\bibinfo{author}{L.~L. Foldy}, \bibinfo{author}{S.~A. Wouthuysen},
\newblock \bibinfo{title}{On the dirac theory of spin 1/2 particles and its non-relativistic limit},
\newblock \bibinfo{journal}{Physical Review} \bibinfo{volume}{78} (\bibinfo{year}{1950}) \bibinfo{pages}{29}. \DOIprefix\doi{https://doi.org/10.1103/PhysRev.78.29}.
\bibitem[{Doost(2016{\natexlab{a}})}]{Doost2015}
\bibinfo{author}{M.~B. Doost},
\newblock \bibinfo{title}{Resonant-state-expansion born approximation with a correct eigen-mode normalisation},
\newblock \bibinfo{journal}{Journal of Optics} \bibinfo{volume}{18} (\bibinfo{year}{2016}{\natexlab{a}}) \bibinfo{pages}{085607}. \DOIprefix\doi{https://doi.org/10.1088/2040-8978/18/8/085607}.
\bibitem[{Doost(2016{\natexlab{b}})}]{Doost2016}
\bibinfo{author}{M.~B. Doost},
\newblock \bibinfo{title}{Resonant-state-expansion born approximation for waveguides with dispersion},
\newblock \bibinfo{journal}{Physical Review A} \bibinfo{volume}{93} (\bibinfo{year}{2016}{\natexlab{b}}) \bibinfo{pages}{023835}. \DOIprefix\doi{https://doi.org/10.1103/PhysRevA.93.023835}.
\bibitem[{Ko and Sambles(1988)}]{Ko1988}
\bibinfo{author}{D.~Y.~K. Ko}, \bibinfo{author}{J.~Sambles},
\newblock \bibinfo{title}{Scattering matrix method for propagation of radiation in stratified media: attenuated total reflection studies of liquid crystals},
\newblock \bibinfo{journal}{JOSA A} \bibinfo{volume}{5} (\bibinfo{year}{1988}) \bibinfo{pages}{1863--1866}. \DOIprefix\doi{https://doi.org/10.1364/JOSAA.5.001863}.
\bibitem[{Tikhodeev et~al.(2002)Tikhodeev, Yablonskii, Muljarov, Gippius, and Ishihara}]{PhysRevB.66.045102}
\bibinfo{author}{S.~G. Tikhodeev}, \bibinfo{author}{A.~L. Yablonskii}, \bibinfo{author}{E.~A. Muljarov}, \bibinfo{author}{N.~A. Gippius}, \bibinfo{author}{T.~Ishihara},
\newblock \bibinfo{title}{Quasiguided modes and optical properties of photonic crystal slabs},
\newblock \bibinfo{journal}{Phys. Rev. B} \bibinfo{volume}{66} (\bibinfo{year}{2002}) \bibinfo{pages}{045102}. \DOIprefix\doi{https://10.1103/PhysRevB.66.045102}.
\bibitem[{Calogeracos and Dombey(1999)}]{Calogeracos2010}
\bibinfo{author}{A.~Calogeracos}, \bibinfo{author}{N.~Dombey},
\newblock \bibinfo{title}{History and physics of the klein paradox},
\newblock \bibinfo{journal}{Contemporary physics} \bibinfo{volume}{40} (\bibinfo{year}{1999}) \bibinfo{pages}{313--321}. \DOIprefix\doi{https://doi.org/10.1080/001075199181387}.
\bibitem[{Dosch et~al.(1971)Dosch, Muller, and Jensen}]{Dosch1971}
\bibinfo{author}{H.~Dosch}, \bibinfo{author}{V.~Muller}, \bibinfo{author}{J.~Jensen},
\newblock \bibinfo{title}{Kleins paradox},
\newblock \bibinfo{journal}{Physica Norvegica} \bibinfo{volume}{5} (\bibinfo{year}{1971}) \bibinfo{pages}{151}.
\bibitem[{Klein(1929)}]{klein1929reflexion}
\bibinfo{author}{O.~Klein},
\newblock \bibinfo{title}{Die reflexion von elektronen an einem potentialsprung nach der relativistischen dynamik von dirac},
\newblock \bibinfo{journal}{Zeitschrift f{\"u}r Physik} \bibinfo{volume}{53} (\bibinfo{year}{1929}) \bibinfo{pages}{157--165}. \DOIprefix\doi{https://doi.org/10.1007/BF01339716}.
\bibitem[{Muljarov et~al.(2011)Muljarov, Langbein, and Zimmermann}]{Muljarov2010}
\bibinfo{author}{E.~A. Muljarov}, \bibinfo{author}{W.~Langbein}, \bibinfo{author}{R.~Zimmermann},
\newblock \bibinfo{title}{Brillouin-wigner perturbation theory in open electromagnetic systems},
\newblock \bibinfo{journal}{EPL (Europhysics Letters)} \bibinfo{volume}{92} (\bibinfo{year}{2011}) \bibinfo{pages}{50010}. \DOIprefix\doi{https://doi.org/10.1209/0295-5075/92/50010}.
\bibitem[{Doost et~al.(2014)Doost, Langbein, and Muljarov}]{Doost2014}
\bibinfo{author}{M.~B. Doost}, \bibinfo{author}{W.~Langbein}, \bibinfo{author}{E.~A. Muljarov},
\newblock \bibinfo{title}{Resonant-state expansion applied to three-dimensional open optical systems},
\newblock \bibinfo{journal}{Physical Review A} \bibinfo{volume}{90} (\bibinfo{year}{2014}) \bibinfo{pages}{013834}. \DOIprefix\doi{https://doi.org/10.1103/PhysRevA.90.013834}.
\bibitem[{Armitage et~al.(2014)Armitage, Doost, Langbein, and Muljarov}]{Armitage2014}
\bibinfo{author}{L.~J. Armitage}, \bibinfo{author}{M.~B. Doost}, \bibinfo{author}{W.~Langbein}, \bibinfo{author}{E.~A. Muljarov},
\newblock \bibinfo{title}{Resonant-state expansion applied to planar waveguides},
\newblock \bibinfo{journal}{Physical Review A} \bibinfo{volume}{89} (\bibinfo{year}{2014}) \bibinfo{pages}{053832}. \DOIprefix\doi{https://doi.org/10.1103/PhysRevA.89.053832}.
\bibitem[{Masir et~al.(2008{\natexlab{a}})Masir, Vasilopoulos, Matulis, and Peeters}]{masir2008direction}
\bibinfo{author}{M.~R. Masir}, \bibinfo{author}{P.~Vasilopoulos}, \bibinfo{author}{A.~Matulis}, \bibinfo{author}{F.~Peeters},
\newblock \bibinfo{title}{Direction-dependent tunneling through nanostructured magnetic barriers in graphene},
\newblock \bibinfo{journal}{Physical Review B} \bibinfo{volume}{77} (\bibinfo{year}{2008}{\natexlab{a}}) \bibinfo{pages}{235443}. \DOIprefix\doi{https://doi.org/10.1103/PhysRevB.77.235443}.
\bibitem[{Masir et~al.(2008{\natexlab{b}})Masir, Vasilopoulos, and Peeters}]{masir2008wavevector}
\bibinfo{author}{M.~R. Masir}, \bibinfo{author}{P.~Vasilopoulos}, \bibinfo{author}{F.~Peeters},
\newblock \bibinfo{title}{Wavevector filtering through single-layer and bilayer graphene with magnetic barrier structures},
\newblock \bibinfo{journal}{Applied Physics Letters} \bibinfo{volume}{93} (\bibinfo{year}{2008}{\natexlab{b}}) \bibinfo{pages}{242103}. \DOIprefix\doi{https://doi.org/10.1063/1.3049600}.

\end{thebibliography}
\end{document}